\documentclass[12pt]{article}   
\pdfoutput=1
\usepackage{geometry,bbm}   
\usepackage{ulem}
\usepackage[toc,page]{appendix}             		
\usepackage{graphicx,cancel}
\usepackage[usenames,dvipsnames]{color}
\usepackage{slashed}					
\usepackage{amssymb,amsmath}
\usepackage{hyperref}
\usepackage{mathtools}
\usepackage{cite}

\usepackage{graphicx}
\usepackage{bm}
\def\bea{\begin{eqnarray}}
\def\eea{\end{eqnarray}}
\usepackage{latexsym}
\usepackage{epsfig,amssymb,euscript}
\usepackage{amsmath}
\usepackage{array,calc,epsfig}

\newcommand{\be}{\begin{equation}}
\newcommand{\ee}{\end{equation}}

\numberwithin{equation}{section}

\begin{document}

\begin{titlepage}

\begin{center}
{\LARGE
{\bf
Complete prepotentials of 5d higher rank theories
}}
\end{center}

\bigskip
\begin{center}
{\large
Francesco Mignosa}
\end{center}

\renewcommand{\thefootnote}{\arabic{footnote}}

\begin{center}
\vspace{0.1cm}
Department of Physics, Technion, Israel Institute of Technology, Haifa, 32000, Israel

\vskip 5pt
{\texttt{francescom@campus.technion.ac.il}}
\end{center}

\vskip 5pt
\noindent
\begin{center} {\bf Abstract} \end{center}
\noindent
We generalize recent methods regarding the construction of complete prepotentials of five dimensional SCFTs to two classes of rank-$N$ theories, i.e. the $X_{1,N}$ and the UV fixed point of the $SU(N+1)_0+2(N+1)$F gauge theories. Their complete prepotentials are consistently constructed starting from their weakly coupled description. The results are verified by employing UV dualities and decoupling of flavors from theories with known complete prepotentials. Applications of these results to the recent investigation of non-supersymmetric conformal field theories from soft supersymmetric breaking deformations are addressed.
\\
\\
\\
\\

\vspace{1.6 cm}
\vfill

\end{titlepage}

\newpage
\tableofcontents

\section{Introduction and summary of results}
\label{intro}
The study of five dimensional gauge theories has recently gained a renewed interest, especially in the context of supersymmetric conformal field theories (SCFT) in diverse dimensions. As the early works on the topic already shown \cite{Seiberg:1996bd,Morrison:1996xf,Intriligator:1997pq,Aharony:1997ju,Aharony:1997bh,DeWolfe:1999hj}, the strong coupling dynamics of $\mathcal{N}=1$ supersymmetric (SUSY) theories in five dimensions can be explored thanks to its large amount of supercharges. In the past years, these theories were explored in detail by many authors, both using field theory techniques and string constructions, see e.g. \cite{Cremonesi:2015lsa,Bergman:2015dpa,Zafrir:2015ftn,Gutperle:2017nwo,Gutperle:2018axv,Hayashi:2018bkd,Hayashi:2018lyv,Hayashi:2019yxj,Xie:2017pfl,Jefferson:2018irk,Bergman:2018hin,Closset:2018bjz,Bhardwaj:2018yhy,Bhardwaj:2018vuu,Bhardwaj:2019jtr,Bhardwaj:2019xeg,Bhardwaj:2020gyu,Closset:2020scj,Bhardwaj:2020ruf,Bhardwaj:2020avz,Apruzzi:2019vpe,Apruzzi:2019opn,Apruzzi:2019enx,Apruzzi:2019kgb,Closset:2019juk,vanBeest:2020kou,Bergman:2020myx,Closset:2020afy,Hubner:2020uvb}. Moreover, very recently, the existence of five dimensional non-supersymmetric conformal field theories was investigated using several techiques, such as the $\epsilon$-expansion, the bootstrap \cite{Fei:2014yja,Nakayama:2014yia,Bae:2014hia,Chester:2014gqa,Li:2016wdp,Arias-Tamargo:2020fow,Li:2020bnb,Giombi:2019upv,Giombi:2020enj}, and soft supersymmety breaking of known superconformal fixed points \cite{BenettiGenolini:2019zth, Bertolini:2021cew, Bertolini:2022osy}. 

In this latter context, the construction of complete prepotentials for low rank theories in \cite{Hayashi:2019jvx} represents a powerful technique to describe their non-perturbative dynamics. Indeed, in many cases superconformal field theories can be mass deformed to infrared free gauge theories whose dynamics, being $\mathcal{N}=1$ supersymmetric, is determined by a function of the Coulomb branch (CB) parameters $\phi_i$, the so-called perturbative prepotential \cite{Seiberg:1996bd, Intriligator:1997pq}. An attempt to classify supersymmetric theories based on this description was attempted in \cite{Intriligator:1997pq}. However, as was pointed out recently in \cite{Jefferson:2017ahm, Jefferson:2018irk}, the perturbative nature of the prepotential is not always able to describe some of the main properties of the corresponding theory and its (possible) UV superconformal fixed point. In particular, the contributions of non-perturbative states to the prepotentials need to be taken into account. Moreover, the perturbative description can be reached from the fixed point via an RG flow triggered by a relevant deformation. This, in general, breaks explicitly part of the global symmetry of the fixed point, so the prepotential of the weakly coupled gauge theory is blind to the full UV global symmetry.

 To take care of these aspects, we can construct a generalized prepotential, which includes all non-perturbative contributions of the hypermultiplets and it is explicitly invariant under the global symmetry of the fixed point. This is denoted in the literature as the complete prepotential of the theory. String theory realizations of five dimensional gauge theories, such as brane web constructions \cite{Aharony:1997bh, Aharony:1997ju}, become particularly useful in this context since they allow to study non-perturbative aspects of these theories systematically.

The complete prepotential is constrained by the global symmetry of the theory, which in many cases can fix it almost completely. When this happens, additional information can be obtained. For example, CS terms for background gauge fields of the global symmetries can be derived from constant terms of the prepotential, which are fixed by the global symmetry of the fixed point. Moreover, the various phases of the theory obtained by deforming the fixed point via a supersymmetric mass deformation are labeled by these topological invariants and some non-trivial properties of the corresponding phase diagram can be revealed. As shown recently in \cite{BenettiGenolini:2019zth, Akhond:2023vlb}, the complete prepotential can give us important information also when a supersymmetry breaking mass deformation is turned on.
\\
\\
In this paper, we use field theory and brane web techniques to construct the complete prepotential of two classes of superconformal field theories: the $X_{1,N}$ theories \cite{Bergman:2018hin} and the UV fixed point of the $SU(N+1)_0+2(N+1)$F theories. This generalizes the analysis in \cite{Hayashi:2019jvx, Akhond:2023vlb} and the results of \cite{Braun:2023fqa}, which were limited to the construction of complete prepotentials of theories of small rank for either the gauge group or the global symmetry.

The rest of the paper is organized as follows. In section \ref{X1N}, we review the pq-web description of the $X_{1,N}$ fixed point and its low energy limit, associated with a linear quiver of $N$ $SU(2)$ nodes. Moreover, we determine the mass parameters associated with the global symmetry of the fixed point, based on the string description of the corresponding global symmetry algebra.

In section \ref{Completeprep}, we review the complete prepotential construction \cite{Hayashi:2019jvx} and we obtain the prepotential of the $X_{1,2}$ theory by decoupling the flavors of the $[4]-SU(2)-SU(2)-[3]$ theory. By virtue of our results, we conjecture a method to construct complete prepotentials of five dimensional superconformal field theories starting from their weakly coupled quiver gauge theory description. 

With this conjecture in mind, in section \ref{X1Nprep} we construct the prepotential of the $X_{1,N}$ theory for generic $N$, verifying its consistency by analyzing the behavior of the complete prepotential for small values of $N$.

In section \ref{SUquiverduality}, we construct the prepotential of $SU(N+1)_0+2(N+1)$F using the techniques of \cite{Hayashi:2019jvx}. Being this theory UV dual to $[2]-SU(2)^N-[2]$, we independently obtain the prepotential of the $X_{1,N}$ theory by decoupling the four additional flavors, giving a further consistency check of our result. 

Finally, in section \ref{conclusion} we conclude with a summary of our results and give some outlook of this work. 

\section{The \boldmath{$X_{1,N}$} theory}
\label{X1N}
To study the $X_{1,N}$ theory it is convenient to set up a framework where the symmetries of the fixed point are manifest. The fixed point can be mass deformed to a low energy description given by a quiver theory 
\begin{equation}
SU(2)-SU(2)-....-SU(2)-SU(2)
\end{equation}
consisting of $N$ $SU(2)$ nodes connected by $N-1$ bifundamentals. When all the mass parameters are turned on, the global symmetry of the quiver is $U(1)_I^N\times U(1)^{N-1}_{BF}$, where $U(1)_I$ stands for the instantonic symmetry associated with each node of the quiver and $U(1)_{BF}$ stands for the global symmetry associated with each bifundamental. We denote by $m_0^{(i)}, \, m_i$ the mass parameters associated with the instantonic and the bifundamental $U(1)$s respectively. Turning on a positive mass for the bifundamentals, the effective coupling associated with each node of the quiver is subject to a finite shift (in the conventions of \cite{Hayashi:2019jvx}) 
\begin{equation}
t_1= \frac{1}{2}m_0^{(1)}-m_1, \,\,\,\, t_i= \frac{1}{2}m_0^{(i)}-m_{i-1}-m_{i} \,\,\, 1<i<N, \,\,\,\, t_N=\frac{1}{2}m_0^{(N)}-m_{N-1}.
\end{equation}
For some specific values of the mass parameters, the global symmetry enhances to a larger group. When all bifundamentals are massless, the symmetry enhances\footnote{In the following, we will not care about the global structure of the enhanced symmetry group.} to a product of non-Abelian groups $U(1)_I^N\times SU(2)_{BF}^{N-1}$. On the other hand, forcing $t_i=0$ for all $i$, the global symmetry enhances to $SU(2)_I^N \times U(1)_{BF}^{N-1}$. Finally, when all the mass parameters are set to zero we end up at the fixed point, where the global symmetry is expected to enhance\footnote{This enhancement was observed from the superconformal index calculations in \cite{Bergman:2013aca} for $X_{1,2}$ and can be also inferred from the isometries of the Higgs branch \cite{Cabrera:2018jxt}.} 
to $SU(2N)$.

\begin{figure}[h!]
\centering
\includegraphics[scale=0.26, trim={1cm, 5cm, 0.3cm, 5cm}, clip]{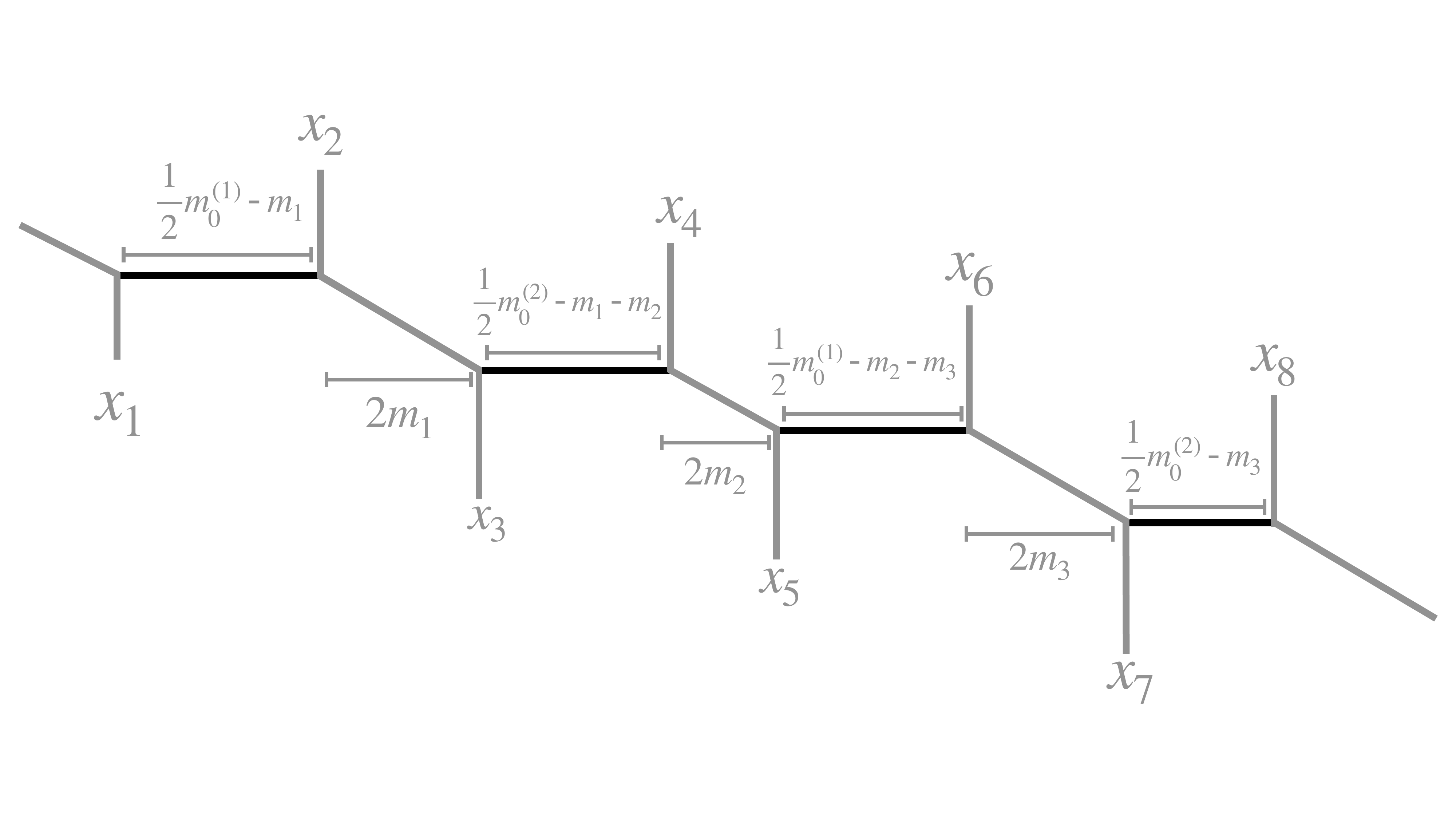}
\caption{Mass parameters of the $X_{1,4}$ theory. Two superimposed D5 branes are depicted in black and describe an $SU(2)$ node.}\label{WeakX1N}
\end{figure}

\begin{figure}[h!]
\centering
\includegraphics[scale=0.26, trim={3cm, 3cm, 0.3cm, 3cm}, clip]{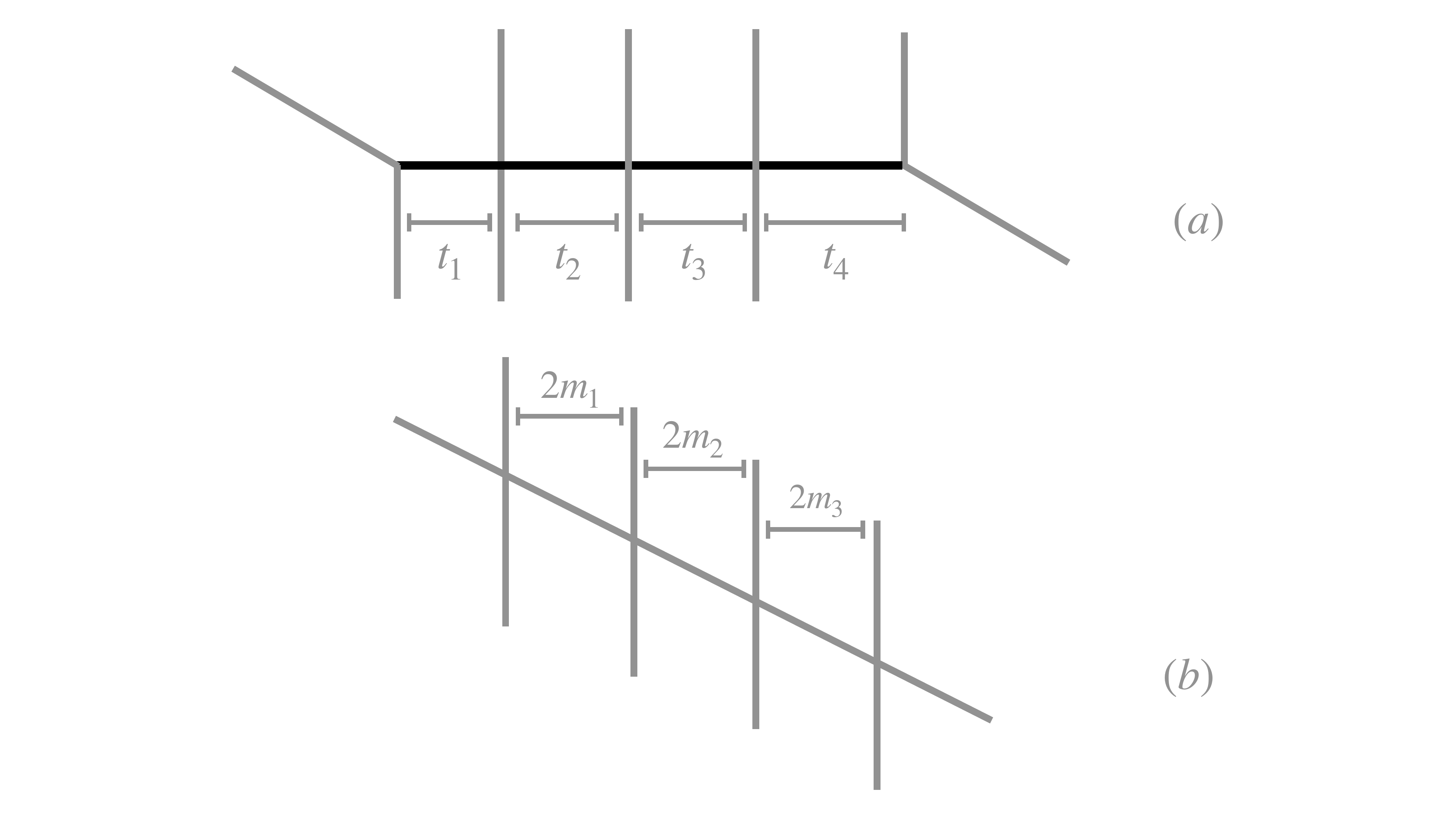}
\caption{Enhancement of bifundamental symmetry (a) and instantonic symmetry (b).}\label{Enhancementofsymmetry}
\end{figure}

The previous reasoning about symmetries is manifest in the brane web language. The pq-web in figure \ref{WeakX1N} hosts the quiver theory at the origin of the Coulomb branch when all mass terms are turned on with a positive value. 
Each couple of (superimposed) D5 branes (drawn in black) host one node of the quiver theory and F1 strings stretching between adjacent couples of D5 branes describe the bifundamental hypermultiplets. When all the bifundamental masses are set to zero, the pq-web reduces to figure \ref{Enhancementofsymmetry}(a). We immediately see that the $SU(2)$ global symmetry of each bifundamental is associated with two NS5 branes lying on the same line. Similarly, tuning the effective couplings $t_i$ to zero, we end up with a series of $E_1$ theories connected by massive bifundamental hypermultiplets, as shown in figure \ref{Enhancementofsymmetry}(b). 
As in the $E_1$ theory \cite{Aharony:1997bh, Bertolini:2021cew}, the enhancement of the instantonic symmetry comes from the alignment of the NS5 branes. 

At the fixed point, the global symmetry enhances to $SU(2N)$. This is manifest if we add 7-branes to the web \cite{DeWolfe:1999hj}. Strings connecting the various $[0,1]$ 7-branes realize an $\mathfrak{su}(2N)$ algebra.\footnote{This can be made more explicit analyzing the monodromy of the 7-branes. From our 7-brane system $X_{[2,-1]}(X_{[0,1]})^N X_{[2,-1]} (X_{[0,1]})^N$, a series of Hanany-Witten transitions reduces the monodromy to $X_{[2,-1]}(X_{[0,1]})^{2N} X_{[2,2N-1]}$, which makes evident the presence of a $\mathfrak{su}(2N)$ subalgebra hosted by the $[0,1]$ 7-branes.} The dynamics of the $2N$ 7-branes is described by a $U(2N)$ gauge theory and the scalars belonging to the corresponding vector multiplet parameterize the transverse position of these branes. However, from the five-dimensional point of view, only the position transverse to the NS5 branes represents a physical parameter \cite{DeWolfe:1999hj}, so the horizontal positions of the 7-branes are associated with the mass parameters of the $SU(2N)$ global symmetry. This is $SU(2N)$ and not $U(2N)$ since a global translation of the whole set of branes represents a redundant parameter for the five-dimensional theory, leaving only the horizontal distances of the 7-branes from the center of mass of the system as physical parameters.  In the following, we will refer to the horizontal position of the $i$-th 7-brane as $x_i$. In particular, we associate with the semi-infinite branes pointing downwards an odd value of $i$, while to the branes pointing upwards, we associate an even value of $i$, as shown in figure \ref{WeakX1N}. Turning on a gauge coupling for each node of the quiver, together with a positive mass for all bifundamentals, we reach the configuration of figure \ref{WeakX1N} and we can describe the position of the 7-branes in terms of the mass parameters of the quiver theory and of the position of the first brane $x_1$ as
\begin{equation}
x_1, \,\,\,x_{n}= x_1+\frac{1}{2}\sum_{j=1}^{[n/2]} m_0^{(j)}+(-1)^{n+1} m_{[n/2]}\,\,\,\forall n<2N, \,\,\ x_{2N}= x_1+\frac{1}{2}\sum_{j=1}^N m_0^{(j)}.
\end{equation}
The position of the center of mass reads
\begin{equation}
x_{\text{COM}}=x_1+\frac{1}{4N}\left(m_0^{(N)}+\sum_{j=1}^{N-1} (2N-2j+1)m_0^{(j)}\right),
\end{equation}
so the relative positions $y_i\equiv x_i-x_{\text{COM}}$ are
\begin{align}\label{MassParameter1}
&y_{1}= -\frac{1}{4N}\sum_{j=1}^{N} (2N-2j+1)m_0^{(j)},\\ \nonumber
&y_{n}= \frac{1}{4N}\left(\sum_{j=1}^{[n/2]} (2j-1)m_0^{(j)}-\sum_{j=[n/2]+1}^{N}(2N-2j+1)m_0^{(j)}\right)+(-1)^{n-1} m_{[n/2]},\\ \nonumber
&y_{2N}= \frac{1}{4N}\sum_{j=1}^{N} (2j-1)m_0^{(j)}\nonumber
\end{align}
and satisfy $\sum_{i=1}^{2N} y_i=0$. For later purposes, it is also convenient to express the bare gauge couplings $m_0^{(n)}$ and the bifundamental mass parameters $m_n$ in terms of the $y_i$s. We see that
\begin{align*}
&m_n=\frac{1}{2}(y_{2n+1}-y_{2n})\,\,\,\forall n=1,...,N-2,\\
&m_1^{(0)}=y_2+y_3-2y_1,\\
&m_n^{(0)}= y_{2n}+y_{2n+1}-y_{2n-1}-y_{2n-2} \,\,\, n=2,...,N-1,\\
&m_N^{(0)}= 2y_{2N}-y_{2N-1}-y_{2N-2}.
\end{align*}
The parameters in eq. (\ref{MassParameter1}) are associated with the $SU(2N)$ global symmetry of the fixed point. In other words, the parameters $m^{(i)}_0, m_i$ of the quiver description mix at the fixed point, and the mass parameters of the $SU(2N)$ global symmetry become linear combinations of the original parameters. In particular, the VEV of the adjoint scalar $Y$ parametrizing the horizontal position of the 7-branes reads
\begin{equation}
\langle Y\rangle = \begin{pmatrix}
y_1 & 0 & 0 & ... & 0\\
0 & y_2 & 0 &....& 0\\
...& ...& ...& ...& ...\\
0 & 0 & 0 &....& y_{2N}
\end{pmatrix}.
\end{equation}
In field theory, $Y$ is identified with the scalar component of the background vector multiplet associated with the global symmetry of the fixed point. 
From this expression, it is clear how the symmetry enhances in some specific limits: tuning the mass of the $i$-th bifundamental to zero is equivalent to making the $(i+1)$-th and the $(i+2)$-th brane coincident, leading to an $SU(2)$ symmetry rotating the corresponding elements of the VEV matrix. Tuning the $i$-th effective coupling $t_i$ to zero makes the $i$-th and the $(i+1)$-th brane coincide, leading to a preserved $SU(2)$ which rotates the corresponding elements of the VEV matrix. Finally, we can consider the supersymmetric deformation analyzed in \cite{Bertolini:2022osy} where all branes with even (resp. odd) labels are forced to lie on the same line and are separated from the group of odd (resp. even) branes by a length $h$. In this case, $Y$ takes the following VEV 
\begin{equation}
\langle Y\rangle = \frac{h}{2}\text{diag}(-1,1,-1,1,...,-1,1).
\end{equation}
This preserves an $SU(N)_L\times SU(N)_R\times U(1)$ global symmetry, as was possible to guess directly from the pq-web since all the even (resp. odd) branes are forced to lie on the same line.

 Finally, the action of the Weyl group $S_{2N}$ of the $SU(2N)$ global symmetry on the mass parameters is easy to determine. This is the permutation group of $2N$ elements acting on the $y_i$ parameters in couples $y_i \leftrightarrow y_j$.
 
Having specified the global symmetry, its mass parameters, and the corresponding action of the Weyl group, we are going to calculate the complete prepotential of the theory in sections \ref{Completeprep} and \ref{X1Nprep}.  
\section{The complete prepotential framework}
\label{Completeprep}
\begin{figure}[h!]
\centering
\includegraphics[scale=0.24, trim={3cm, 3cm, 0.3cm, 3cm}, clip]{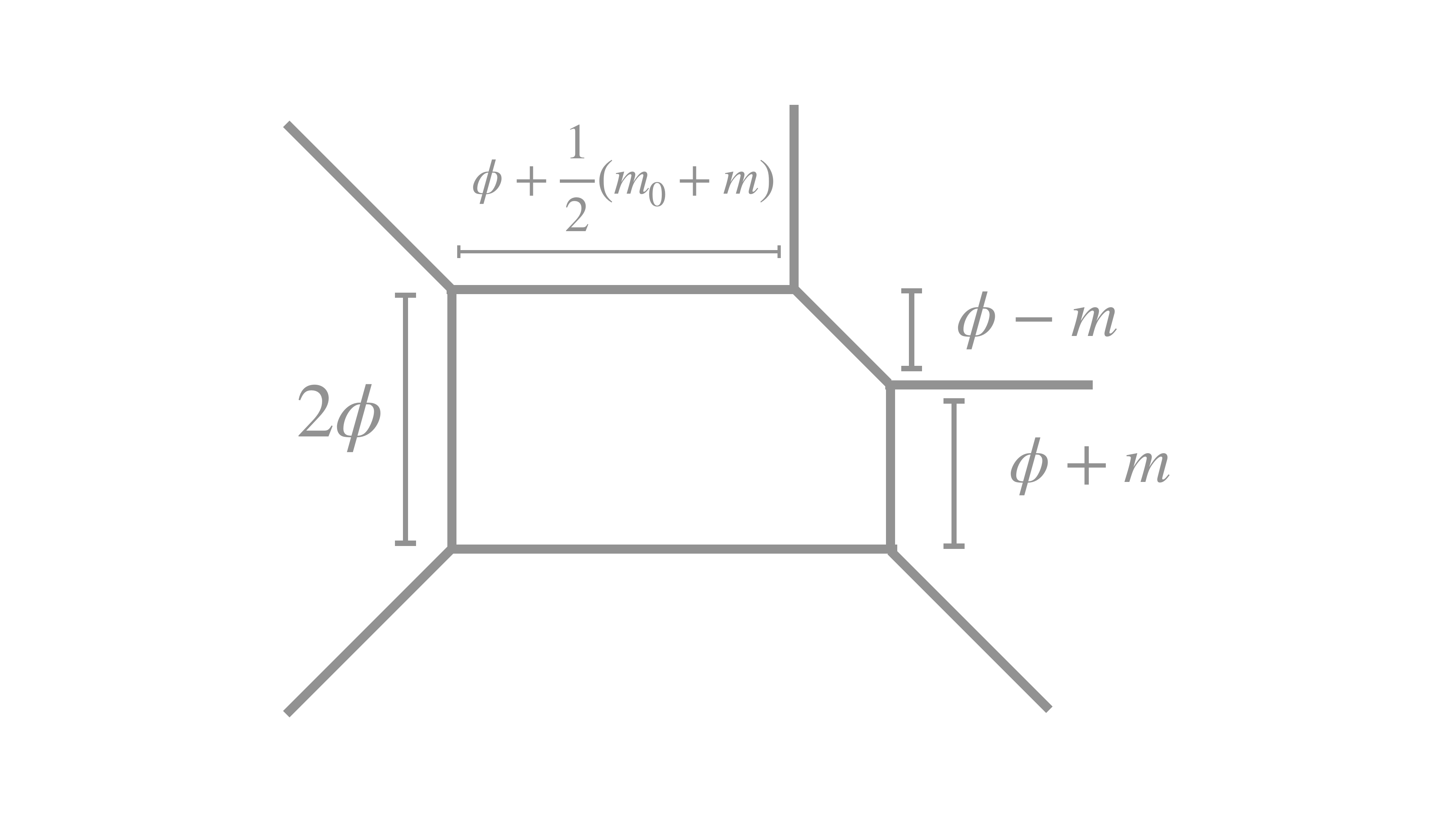}
\caption{$SU(2)$+1F pq-web.}\label{SU(2)1F}
\end{figure}

Given a 5d $\mathcal{N}=1$ supersymmetric gauge theory, its perturbative dynamics is determined by the prepotential function $\mathcal{F}(\phi_i)$ \cite{Intriligator:1997pq}. This is one-loop exact and at most cubic in the Coulomb branch (CB) parameters $\phi_i, \,\, i=1,...,N$ where $N$ is the rank of the theory. It is also manifestly symmetric under the perturbative global symmetry of the theory, but it is blind to the non-perturbative effects that are responsible for the enhancement of the symmetry at the fixed point. As a consequence, many strong coupling aspects of these theories remain inaccessible if we limit ourselves to the study of their perturbative prepotential. On the other hand, recent attempts to classify field theories admitting a five dimensional UV completion to a superconformal field theory \cite{Jefferson:2017ahm, Jefferson:2018irk} suggested the possibility of improving the prepotential to take care of non-perturbative effects \cite{Hayashi:2019jvx}, taking advantage of the pq-web description of the theory. The corresponding prepotential is denoted in the literature as the complete prepotential of the theory. Let us briefly summarize the procedure to construct it from the pq-web:
\begin{itemize}
\item Firstly, we need to identify the CFT phase of the theory: in this phase, all the mass parameters are smaller than the CB ones and all the hypermultiplets of the theory are massive on the CB. Setting the mass parameters to zero, we smoothly reach a point on the CB of the CFT. For this reason, the prepotential of this phase should be manifestly invariant under the global symmetry of the fixed point. 
\item To identify the CFT phase of the theory, we can start analyzing the low energy gauge theory description via its pq-web construction. The CFT phase can be reached by flopping some edges of this web. In field theory, the flop corresponds to a change in the sign of the mass $M_f$ of a corresponding hypermultiplet. This operation modifies the prepotential, which acquires additional terms proportional to the cube of $M_f$. When we reach the CFT phase, the prepotential is obtained by looking at the tension of the D3 branes wrapping the various faces of the pq-web \cite{Aharony:1997bh}. The corresponding effective couplings, obtained as derivatives of the monopole tensions, are, by dimensional analysis, linear combinations of the CB and mass parameters. Being the CFT point invariant under the full global symmetry, these couplings should be invariant under the action of its Weyl group. This transforms both the CB and the mass parameters in general, since self-dualities of the low energy field theory represent a symmetry from the CFT point of view \cite{Mitev:2014jza}. As a consequence, the effective couplings are necessarily linear combinations of the CB and mass parameters that are left invariant by the action of the symmetry. These combinations, defined as the invariant CB parameters of the theory $\Phi_i$, can be interpreted as the natural coordinates of the CFT Coulomb branch.
\item Having obtained the expression for $\Phi_i$, the prepotential in this phase can be described as a function of the invariant CB parameters and the mass parameters. This is invariant under the full global symmetry group and describes the dynamics of the theory on the CB around the fixed point. For this reason, this is denoted in the literature as the CFT prepotential $\mathcal{F}_{\text{CFT}}$.
\item All other phases of the theory can be reached from the CFT one by flipping the sign of the mass $M_f$ of some hypermultiplet. In the operation, some edge of the corresponding pq-web gets flopped. A generic BPS hypermultiplet of charge $q^i$ under the $i$-th Cartan $U(1)$ of the gauge group and weight $w^f_a$ under the global symmetry of the fixed point contributes to the prepotential through its mass 
\begin{equation}\label{massofhyper}
M_f=q^i \phi_i +w^f_a m_a
\end{equation}
via a cubic term of the form
\begin{equation}
\mathcal{F}= \frac{1}{6}[|M_f|]^3,
\end{equation}
where $[|x|]= \frac{1}{2}(x-|x|)$ and $m_a, \,\, a=1,...,g$ label all the mass parameters associated with the global symmetry of rank $g$ enjoyed by the fixed point. In the CFT phase, these contributions vanish being $M_f>0$ and the prepotential reduces to the CFT one. However, if we move from the CFT phase to another phase of the theory, some hypermultiplet changes the sign of its mass and the prepotential of the theory is shifted by the following amount
\begin{equation}
\Delta \mathcal{F}= \frac{1}{6}M_f^3. 
\end{equation}
  The complete prepotential $\mathcal{F}_{\text{compl.}}$ can be then obtained as a sum of the original $\mathcal{F}_{\text{CFT}}$ prepotential and of all possible flops $\frac{1}{6}\sum_f [|M_f|]^3$.
\end{itemize}
As an example to illustrate the previous procedure, we can calculate the prepotential of $E_2$ theory. This is the UV completion of $SU(2)$ SYM with a single flavor. Its perturbative global symmetry is $SO(2)_\text{F}\times U(1)_I$, where the $SO(2)_\text{F}$ group is associated with the single flavor and to the mass parameter $m$, and $U(1)_I$ with the instantonic symmetry and the mass parameter $m_0\equiv \frac{1}{g^2}$. Its pq-web realization is shown in figure \ref{SU(2)1F}. The $SO(2)_\text{F}\times U(1)_I$ global symmetry enhances at the fixed point to $E_2=U(1)\times SU(2)$. $S$-duality \cite{Hayashi:2019jvx} allows us to identify the mass parameters associated with the $SU(2)$ and the $U(1)$ groups respectively
\begin{equation}
x=\frac{1}{4}m_0-\frac{1}{4}m, \,\,\,y=-\frac{1}{4}m_0-\frac{7}{4}m,
\end{equation}
where $x,y$ transform as $x\rightarrow -x, \,\, y\rightarrow y$ under the Weyl group $\mathbb{Z}_2$ of the $SU(2)$ factor.\footnote{Note that the expression of $x$ in terms of the low energy mass parameters $m_0,m$ could also be inferred directly by looking at the corresponding 7-brane realization of the global symmetry algebra. In particular, this is associated with the distance between the two $[1,-1]$ 7-branes.}
In the CFT phase $\phi\gg |m|, |m_0|$, the pq-web reduces to figure \ref{SU(2)1F}. 
The monopole tension $T$ is proportional to the area of the face
\begin{equation}
T=\frac{7}{2}\phi^2+m_0\phi-\frac{m^2}{2}
\end{equation}
and the effective coupling reduces to
\begin{equation}
\frac{\partial T}{\partial \phi}=g^{-2}_{eff}= 7\phi +m_0.
\end{equation}
We can then introduce an invariant CB parameter $\Phi$ as\footnote{The invariance of this parameter is manifest being it proportional to the effective coupling itself.}
\begin{equation}
\Phi\equiv\phi+\frac{1}{7}m_0= \frac{1}{7} g_{eff}^{-2}
\end{equation}
and write the CFT prepotential as
\begin{equation}
\mathcal{F}_{\text{CFT}}= \frac{7}{6}\Phi^3-\frac{1}{14}m_0^2\Phi-\frac{m^2}{2} \Phi.
\end{equation}
The hypermultiplets of the theory are realized\footnote{Notice that not all hypermultiplets can be realized, in the absence of 7-branes, as BPS configurations of strings ending on 5-branes, see \cite{Aharony:1997bh}.} in string theory as F1 strings connecting the semi-infinite D5 brane with the finite dimensional ones and as D1 strings connecting the external and internal NS5 branes. These have masses $\phi \pm m, \,\, \phi+\frac{1}{2}m_0+\frac{m}{2}$ respectively, as can be deduced from figure \ref{SU(2)1F}. This leads to the following hypermultiplet contribution
\begin{equation}
\mathcal{F}_{\text{hyper}}=\frac{1}{6}[|\phi-m|]^3+\frac{1}{6}[|\phi+m|]^3+\frac{1}{6}\left[\left|\phi+\frac{1}{2}m_0+\frac{m}{2}\right|\right]^3. 
\end{equation}
Adding this contribution to the CFT prepotential we obtain the complete prepotential of the theory
\begin{equation}
\mathcal{F}_{\text{compl.}}= \frac{7}{6}\Phi^3 -\left(x^2+\frac{1}{7}y^2\right)\Phi+\frac{1}{6}\left[\left|\Phi+\frac{4}{7}y\right|\right]^3+\frac{1}{6}\left[\left|\Phi\pm x-\frac{3}{7}y\right|\right]^3,
\end{equation}
with $\Phi=\phi+\frac{1}{2}x-\frac{y}{14}$. The representations of the hypermultiplets under the $E_2$ group can be read directly from their mass, see eq. (\ref{massofhyper}). In particular, the first particle is a singlet of $SU(2)$ with charge $4/7$ under $U(1)$, while the second is a doublet under $SU(2)$ with charge $-3/7$ under the remaining $U(1)$. \\
\newline
Following the method explained above, we will calculate the complete prepotential of the $X_{1,N}$ theory in section \ref{X1Nprep}. However, before doing that, we will first obtain independently the prepotential in the $N=2$ case: this will serve as a toolkit when we will perform the calculation for the generic $N$ case. 
\subsection{$X_{1,2}$ prepotential from decoupling}
\label{X12fromSp2}
\begin{figure}[h!]
\centering
\includegraphics[scale=0.24, trim={0.3cm, 1cm, 0.3cm, 1cm}, clip]{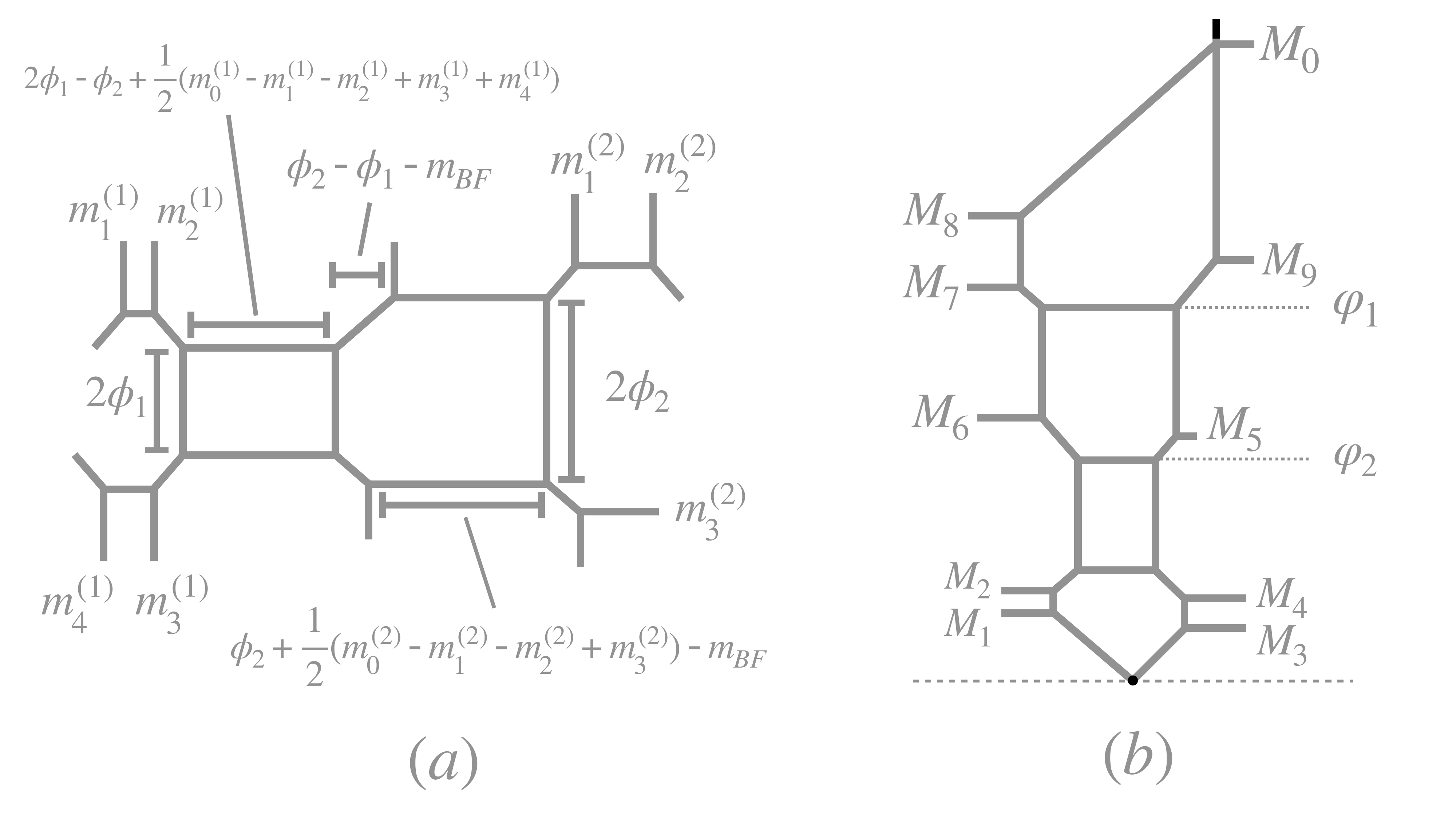}
\caption{Duality between $[4]-SU(2)-SU(2)-[3]$ and $Sp(2)$+9F.}\label{SU2SU2F92}
\end{figure}

We can obtain the complete prepotential of the $X_{1,2}$ theory by decoupling all the flavors of the $[4]-SU(2)-SU(2)-[3]$ theory, for which the complete prepotential was obtained in \cite{Hayashi:2019jvx}. In particular, the theory is UV dual to $Sp(2)$+9F, as shown in figure \ref{SU2SU2F92}. At the CFT point the $Spin(18)\times U(1)_I$ global symmetry of the symplectic theory enhances to $SO(20)$, which mass parameters $M_0, M_i, \,\,\, i=1,...,9$ are respectively the inverse gauge coupling square and the nine flavor masses of the $Sp(2)$ theory. 
The complete prepotential can be written in these variables as
\begin{align*}
&6\mathcal{F}= \tilde{\varphi}_1^3-2\tilde{\varphi}_2^3+ 6\tilde{\varphi}_1\tilde{\varphi}_2^2-3\sum_{i=0}^9 M_i^2 (\tilde{\varphi}_1+\tilde{\varphi}_2)+\sum_{0\leq i \leq j \leq 9}[|\tilde{\varphi}_1 \pm M_i\pm M_j|]^3+\sum_{i=0}^9[|\tilde{\varphi}_2\pm M_i|]^3+ \\
&\sum_{\{s_i=\pm 1, \text{odd}+\}}[|\tilde{\varphi}_1+\tilde{\varphi}_2+\frac{1}{2} \sum_{i=0}^9 s_i M_i|]^3+\sum_{0\leq i_1<i_2<i_3<i_4<i_5\leq 9} [|2\tilde{\varphi}_1+\tilde{\varphi}_2\pm M_{i_1}\pm M_{i_2}\pm M_{i_3}\pm M_{i_4}\pm M_{i_5}|]^3,
\end{align*}
where the expression $[|a\pm b\pm c|]^3$ stands for the sum of four terms 
\begin{equation}
[|a+b+c|]^3+[|a-b+c|]^3+[|a-b-c|]^3+[|a+b-c|]^3,
\end{equation}
and the invariant CB parameters read
\begin{align*}
\tilde{\varphi}_1= \varphi_1+M_0, \,\,\,\, \tilde{\varphi}_2=\varphi_2.
\end{align*}
Here $\sum_{\{s_i=\pm 1, \, \text{odd}+\}}$ denotes the sum over all possible combinations of $s_i=\pm 1$ such that the total number of $s_i$ taking value $+1$ is odd. The UV duality, which manifests in the string theory construction as $S$-duality, relates these variables to the quiver ones via the following relations for the mass parameters
\begin{align}
M_1=&\frac{1}{2}(-m_1^{(1)}+m_2^{(1)}-m_3^{(1)}-m_4^{(1)}), \,\,\,\, M_2= \frac{1}{2}(m_1^{(1)}-m_2^{(1)}-m_3^{(1)}-m_4^{(1)}),\label{Dualityrelations}\\ \nonumber
M_3=&\frac{1}{2}(m_1^{(1)}+m_2^{(1)}-m_3^{(1)}+m_4^{(1)}) , \,\,\,\, M_4=\frac{1}{2}(m_1^{(1)}+m_2^{(1)}+m_3^{(1)}-m_4^{(1)}),\\ \nonumber
M_5=&\frac{1}{2}m_0^{(1)}+m_{BF}, \,\,\, M_6= \frac{1}{2} m_0^{(1)}-m_{BF}, \,\,\, M_7=\frac{1}{2}(m_0^{(1)}+m_0^{(2)}-m_1^{(2)}+m_2^{(2)}+m_3^{(2)}), \\\nonumber
M_8=&\frac{1}{2}(m_0^{(1)}+m_0^{(2)}+m_1^{(2)}-m_2^{(2)}+m_3^{(2)}),\,\,\,M_9=\frac{1}{2}(m_0^{(1)}+m_0^{(2)}-m_1^{(2)}-m_2^{(2)}-m_3^{(2)}),\\\nonumber
M_0=&\frac{1}{2}(m_0^{(1)}+m_0^{(2)}+m_1^{(2)}+m_2^{(2)}-m_3^{(2)})\nonumber
\end{align}
and
\begin{align*}
\varphi_1=& \phi_2+\frac{1}{2}(m_0^{(1)}+m_0^{(2)}-m_1^{(2)}-m_2^{(2)}+m_3^{(2)}),\\
\varphi_2=& \phi_1-\phi_2+\frac{1}{2}m_0^{(1)},
\end{align*}
for the CB parameters.

The prepotential is written in a specific Weyl chamber of the $Sp(2)$ theory \cite{Hayashi:2019jvx} specified by the relations $\varphi_1\geq \varphi_2\geq 0$. This constrains, through the duality relations, the CB parameters of the quiver theory. Note that, in this case, the CFT phase is separated from the perturbative one of the $Sp(2)$ theory by a flop of an $Sp(2)$ instanton of mass $\varphi_2-M_0$ \cite{Hayashi:2019jvx}.\\
\\
To obtain the prepotential of $X_{1,2}$, we can start from the $[4]-SU(2)-SU(2)-[3]$ and decouple the flavors one at a time. Looking at the pq-web, this operation consists of sending the mass of the flavor to infinity while keeping fixed the lengths of the remaining pq-web. In our case, this requires sending some other mass (or CB) parameter to infinity. Let us consider the decoupling of the flavor of the second node of the quiver with mass $m_1^{(2)}$. To keep the remaining lengths of the web finite $m_0^{(2)}$ should be sent to infinity also, while $m_0^{(2)}-m_1^{(2)}$ is kept fixed. Through the duality relations in eq. (\ref{Dualityrelations}), the decoupling induces the following limit
\begin{equation}
M_8\rightarrow +\infty,\,\, M_0\rightarrow +\infty, \,\,\, M_0-M_8=\text{fixed}.
\end{equation}
This shows that the decoupling of one flavor in the quiver theory translates into the decoupling of a flavor in the symplectic theory. The theory reduces to $Sp(2)$+8F, which is UV dual to $[4]-SU(2)-SU(2)-[2]$. At the fixed point, the global symmetry is $SO(16)\times SU(2)_I$.

Similarly, decoupling another flavor associated with the second node of the quiver translates into decoupling one flavor of $Sp(2)$ theory, establishing the UV duality $Sp(2)+7\text{F}\leftrightarrow [4]-SU(2)-SU(2)-[1]$. In this case, the global symmetry $SO(14)_F\times U(1)_I$ of the $Sp(2)$ theory does not enhance at the fixed point, so the invariant CB parameters coincide with the weak coupling ones. After renaming the new gauge coupling as $M_0$ and $M_9$ as $M_7$, the complete prepotential reduces to
\begin{equation}
6\mathcal{F}_{Sp(2)+7\text{F}}= 6\mathcal{F}_{\text{CFT}}+6\mathcal{F}_{\text{singlet}}+6\mathcal{F}_{F}+6\mathcal{F}_{C}+6\mathcal{F}_{\text{AS}},
\end{equation}
where
\begin{align*}
&6\mathcal{F}_{\text{CFT}}= 3 \varphi_1^3+6\varphi_1\varphi_2^2+\varphi_2^3+3M_0(\varphi_1^2+\varphi_2^2)-3\sum_{i=1}^7 M_i^2(\varphi_1+\varphi_2),\\
&6\mathcal{F}_{\text{singlet}}= [|\varphi_1\pm M_0|]^3,\\
&6\mathcal{F}_F=\sum_{I=1}^2 \sum_{i=1}^7 [|\varphi_I\pm M_i|]^3,\\
&6\mathcal{F}_C=\sum_{\{s_i=\pm 1, \text{odd}+\}} [|\varphi_1+\varphi_2+\frac{1}{2}M_0+\frac{1}{2} \sum_{i=1}^7 s_i M_i|]^3,\\
&6\mathcal{F}_{\text{AS}}= \sum_{1\leq i<j\leq 7}[|2\varphi_1+\varphi_2+M_0\pm M_i\pm M_j|]^3
\end{align*}
and the duality relations read
\begin{align*}
M_1=&\frac{1}{2}(-m_1^{(1)}+m_2^{(1)}-m_3^{(1)}-m_4^{(1)}), \,\,\,\, M_2= \frac{1}{2}(m_1^{(1)}-m_2^{(1)}-m_3^{(1)}-m_4^{(1)}),\\
M_3=&\frac{1}{2}(m_1^{(1)}+m_2^{(1)}-m_3^{(1)}+m_4^{(1)}) , \,\,\,\, M_4=\frac{1}{2}(m_1^{(1)}+m_2^{(1)}+m_3^{(1)}-m_4^{(1)}),\\
M_5=&\frac{1}{2}m_0^{(1)}+m_{BF}, \,\,\, M_6= \frac{1}{2} m_0^{(1)}-m_{BF}\\
M_7=&\frac{1}{2}(m_0^{(1)}+m_0^{(2)}-m_3^{(2)}),\,\,\,\,M_0=\frac{1}{2}(-3m_3^{(2)}-m_0^{(1)}-m_0^{(2)}),\\
\varphi_1=& \phi_2+\frac{1}{2}(m_0^{(1)}+m_0^{(2)}+m_3^{(2)}),\\
\varphi_2=& \phi_1-\phi_2+\frac{1}{2}m_0^{(1)}.
\end{align*}
 From now on, decoupling a flavor in the quiver theory will not correspond to decoupling a flavor in the $Sp(2)$ theory. However, it is nevertheless convenient to work out the decoupling limit in terms of the $Sp(2)$ variables, since the global symmetry of the fixed point is manifest in this parametrization. Let us decouple the last flavor of the second node
 \begin{equation}
 m_3^{(2)}\rightarrow +\infty, \,\, m_0^{(2)}\rightarrow+\infty, \,\, m_0^{(2)}-m_3^{(2)}=\text{fixed}
 \end{equation}
 which, in terms of the $Sp(2)$ parameters boils down to
 \begin{equation}
 \varphi_1\rightarrow+\infty,\,\, M_0\rightarrow -\infty, \,\,\, \hat{\varphi}_1\equiv \varphi_1+\frac{1}{2}M_0=\text{fixed}.
 \end{equation}
 This reduces the prepotential to (after redefining $\hat{\varphi}_1\rightarrow \varphi_1$)
 \begin{align}\label{[4]-SU(2)-SU(2) prep}
&6\mathcal{F}_{[4]-SU(2)-SU(2)}= 4\varphi_1^3+6\varphi_1\varphi_2^2+\varphi_2^3 -3\sum_{i=1}^7 M_i^2(\varphi_1+\varphi_2)+\\
&+\sum_{i=1}^7[|\varphi_2\pm M_i|]^3+\sum_{\{s_i=\pm 1, \, \text{odd}+\}}[|\varphi_1+\varphi_2+\frac{1}{2}\sum_{i=1}^7 s_i M_i|]^3+\nonumber\\
&+\sum_{i\leq i<j\leq 7}[|2\varphi_1+\varphi_2\pm M_i\pm M_j|]^3\nonumber
\end{align}
with parameters 
\begin{align*}
M_1=&\frac{1}{2}(-m_1^{(1)}+m_2^{(1)}-m_3^{(1)}-m_4^{(1)}), \,\,\,\, M_2= \frac{1}{2}(m_1^{(1)}-m_2^{(1)}-m_3^{(1)}-m_4^{(1)}),\\
M_3=&\frac{1}{2}(m_1^{(1)}+m_2^{(1)}-m_3^{(1)}+m_4^{(1)}), \,\,\,\, M_4=\frac{1}{2}(m_1^{(1)}+m_2^{(1)}+m_3^{(1)}-m_4^{(1)}),\\
M_5=&\frac{1}{2}m_0^{(1)}+m_{BF}, \,\,\, M_6= \frac{1}{2} m_0^{(1)}-m_{BF}, \,\,\,\,
M_7=\frac{1}{2}(m_0^{(1)}+m_0^{(2)}),\\
&\varphi_1= \phi_2+\frac{1}{4}m_0^{(1)}+\frac{1}{4}m_0^{(2)},\\
&\varphi_2= \phi_1-\phi_2+\frac{1}{2}m_0^{(1)}.
\end{align*}
The theory manifests an $SO(14)$ global symmetry, since the hypermultiplets in eq. \eqref{[4]-SU(2)-SU(2) prep} sit in the $\bold{14}$, the $\bold{\overline{64}}$ and $\bold{91}$ representations of $SO(14)$ respectively. The $U(1)_I$ instantonic symmetry of the symplectic theory associated with the $M_0$ mass parameter decouples in the operation and the resulting theory enjoys an $SO(14)$ global symmetry.\\
\\
If we now try to decouple another flavor, the whole set of $Sp(2)$ mass parameters diverges. Indeed, sending 
\begin{equation}
m_1^{(1)}\rightarrow +\infty, \,\, m_0^{(1)}\rightarrow+\infty, \,\, m_0^{(1)}-m_1^{(1)}=\text{fixed}
\end{equation}
the mass parameters diverge as
\begin{equation}
M_2,M_3,M_4,M_5,M_6,M_7\rightarrow +\infty, \,\,\, M_1\rightarrow-\infty, \,\,\, \varphi_1\rightarrow +\infty, \,\,\, \varphi_2\rightarrow +\infty.
\end{equation}
To regularize the prepotential, we can in principle use any linear combination of mass parameters to subtract the divergence and redefine the new (finite) mass parameters. In order to choose the correct subtraction, we rely on the symmetries that should be preserved after the decoupling. Sending $m_1^{(1)}\rightarrow+\infty$, the global symmetry of the theory $G\equiv SO(14)$ is expected to be reduced to a subgroup $G'\subset G$. The subtraction of divergences is then dictated by the embedding of $G'$ into $G$. This becomes manifest when we look at how the mass parameters diverge in terms of $m_0^{(1)}$: $M_i$ with $i\neq 1$ diverge as $\sim \frac{1}{2} m_0^{(1)}$, while $M_1$ diverges as $\sim -\frac{1}{2}m_0^{(1)}$. We can subtract divergences by maintaining the maximal amount of symmetry by redefining
\begin{align*}
&y_1=-M_1+z, \,\,\, y_i=M_i+z\,\,\,\forall i=2,...,7,\\
&\Phi_1=\varphi_1+z/2, \,\,\,\, \Phi_2=\varphi_2+z
\end{align*}
with $z\equiv \frac{1}{7}(M_1-\sum_{j=2}^7 M_j)$ and $\sum_i y_i=0$. The complete prepotential reads
\begin{align*}
&6\mathcal{F}_{[3]-SU(2)-SU(2)}= 5\Phi_1^3+9\Phi_1\Phi_2^2+2\Phi_2^3+3\Phi_1^2\Phi_2-3\sum_i y_i^2(\Phi_1+\Phi_2)+\\
&+\sum_{i=1}^7[|\Phi_2- y_i|]^3+\sum_{1\leq i<j \leq 7}[|2\Phi_1+\Phi_2-y_i-y_j|]^3+\sum_{1\leq i<j \leq 7}[|\Phi_1+\Phi_2+y_i+y_j|]^3
\end{align*}
with CB parameters
\begin{equation}
\Phi_1=\phi_2+\frac{1}{14}(2m_0^{(1)}+3m_0^{(2)}),\,\,\,\Phi_2=\phi_1-\phi_2+\frac{1}{14}(4m_0^{(1)}-m_0^{(2)}).
\end{equation}
We see that the hypermultiplets organize in the $\bold{7}$, $\bold{\overline{21}}$ and $\bold{21}$ representations of an $SU(7)$ global symmetry.\footnote{Note that the masses of these hypermultiplets, being finite under the decoupling process, do not depend on the choice adopted to subtract the divergences. This is not true, instead, for the CFT prepotential.} In other words, only an $SU(7)\times U(1)$ subgroup of $SO(14)$ remains preserved after the decoupling, where the $U(1)$ factor is associated with the $z$ parameter and the $SU(7)$ subgroup with the $y_i$ parameters. Sending $z$ to infinity reduces the global symmetry of the theory to $SU(7)$ with mass parameters
\begin{align*}
&y_1=\frac{1}{14}(-3m_0^{(1)}-m_0^{(2)}-7m_2^{(1)}+7m_3^{(1)}+7m_4^{(1)}), \\
&y_2=\frac{1}{14}(-3m_0^{(1)}-m_0^{(2)}-7m_2^{(1)}-7m_3^{(1)}-7m_4^{(1)}),\\
&y_3=\frac{1}{14}(-3m_0^{(1)}-m_0^{(2)}+7m_2^{(1)}-7m_3^{(1)}+7m_4^{(1)}),\\
&y_4=\frac{1}{14}(-3m_0^{(1)}-m_0^{(2)}+7m_2^{(1)}+7m_3^{(1)}-7m_4^{(1)}),\\
&y_5=\frac{1}{14}(4m_0^{(1)}-m_0^{(2)})+m_{BF}, \\
&y_6=\frac{1}{14}(4m_0^{(1)}-m_0^{(2)})-m_{BF},\\
&y_7=\frac{1}{7}(2m_0^{(1)}+3m_0^{(2)}).
\end{align*}
Similarly, we can proceed decoupling all the flavors one by one, regularizing them in a way that respects the maximum amount of symmetry. Decoupling one flavor we reach $[2]-SU(2)-SU(2)$ with $SU(5)\times SU(2)$ global symmetry\footnote{The symmetry group can be analogously obtained by looking at the 7-brane monodromy of the corresponding pq-web as in \cite{Mitev:2014jza}.} and decoupling two of them we reach $[1]-SU(2)-SU(2)$ with an $SU(4)\times U(1)$ global symmetry. Decoupling the last flavor with a negative mass, we reach the $SU(2)_\pi-SU(2)_0$ theory, which was shown to enjoy an $SU(3)\times U(1)$ global symmetry \cite{Bergman:2013aca}, while decoupling with a positive mass we reach the $SU(2)-SU(2)$ theory, with global symmetry $SU(4)$. The prepotential reads
\begin{align}\label{X12prepot}
&6\mathcal{F}_{SU(2)-SU(2)}= 8\Phi_1^3+18\Phi_1\Phi_2^2+5\Phi_2^3+12\Phi_1^2\Phi_2-3\sum_{i=1}^4 y_i^2(\Phi_1+\Phi_2)\\
&+\sum_{i=1}^4[|\Phi_2- y_i|]^3+\sum_{1\leq i<j \leq 4}[|2\Phi_1+\Phi_2-y_i-y_j|]^3\nonumber
\end{align}
with mass parameters
\begin{align*}
&y_1=\frac{1}{8}(-m_0^{(2)}-3m_0^{(1)}), \,\,\, y_2=\frac{1}{8}(-m_0^{(2)}+m_0^{(1)})-m_{BF}, \\
&y_3=\frac{1}{8}(-m_0^{(2)}+m_0^{(1)})+m_{BF}, \,\,\,\, y_4=\frac{1}{8}(3m_0^{(2)}+m_0^{(1)})
\end{align*}
and invariant CB parameters
\begin{equation}
\Phi_1=\phi_2+\frac{1}{16}(3m_0^{(2)}+m_0^{(1)}), \,\,\, \Phi_2=\frac{1}{8}(-m_0^{(2)}+m_0^{(1)})+\phi_1-\phi_2.
\end{equation}
The hypermultiplets in eq. (\ref{X12prepot}) are in the $\bold{4}$ and $\bold{6}$ representation of the $SU(4)$ global symmetry, as can be inferred from their masses. As anticipated in section \ref{X1N}, the mass parameters of the $SU(4)$ global symmetry match the general form in eq. (\ref{MassParameter1}) for the $X_{1,N}$ theory, giving a consistency check of the validity of our decoupling procedure. It is convenient to rearrange the invariant CB parameter more symmetrically defining
\begin{equation}
\bar{\Phi}_1\equiv\Phi_1+\Phi_2= \phi_1+\frac{1}{16}(3m_0^{(1)}+m_0^{(2)}), \,\,\, \bar{\Phi}_2=\Phi_1.
\end{equation}
After redefining $\bar{\Phi}_i \rightarrow \Phi_i$, the prepotential can be rewritten as
\begin{align*}
&6\mathcal{F}_{SU(2)-SU(2)}= 5\Phi_1^3-9\Phi_1\Phi_2^2+9\Phi_2^3+3\Phi_1^2\Phi_2-3\sum_{i=1}^4 y_i^2\Phi_1\\
&+\sum_{i=1}^4[|\Phi_1-\Phi_2-y_i|]^3+\sum_{1\leq i<j \leq 4}[|\Phi_1+\Phi_2-y_i-y_j|]^3.
\end{align*}
Several comments are now in order.

The prepotential matches the perturbative prepotential $\mathcal{F}_{\text{weak}}$ for the $SU(2)-SU(2)$ quiver theory
\begin{equation}
\mathcal{F}_{\text{weak}}= \frac{4}{3}(\phi_1^3+\phi_2^3)+\frac{1}{2}m_0^{(1)}\phi_1^2+\frac{1}{2}m_0^{(2)}\phi_2^2-\frac{1}{12}|\phi_1\pm \phi_2\pm m_{BF}|^3
\end{equation}
in the weak coupling limit $m_0^{(1)}, m_0^{(2)}\gg \phi_i, |m_{BF}|$. In particular, the hypermultiplet contribution of the bifundamental fields 
\begin{equation}
6\mathcal{F}_{\text{hyper}}\subset [|\phi_1+\phi_2\pm m_{BF}|]^3+[|\phi_1-\phi_2\pm m_{BF}|]^3
\end{equation}
indicates that in the CFT phase of the quiver theory, the CB parameters are ordered as 
\begin{equation}\label{orderingcondition}
\phi_1\geq \phi_2.
\end{equation} 
This is reminiscent of the Weyl chamber condition for the $[4]-SU(2)-SU(2)-[3]$ case and it has an analog in the $SU(N)$ theories. Usually for these theories, we describe the CB using $N$ CB parameters $a_i, \,\, i=1,...,N$ which are not linearly independent $\sum_i a_i=0$. In this parametrization, we can select a Weyl chamber by ordering the parameters as $a_1\geq a_2\geq...\geq a_{N-1}\geq 0\geq a_N$. However, we can also parametrize the CB in the Dynkin basis using $N-1$ independent coordinates $\phi_i$ related to $a_i$ as
\begin{equation}
a_1=\phi_1, \, a_i=\phi_i-\phi_{i-1} \,\,\, i=2,...,N-1, \, a_N=-\phi_{N-1}.
\end{equation}
In this parametrization, the Weyl chamber is described through the conditions $\phi_i\geq \phi_{i-1}, \,\, \phi_i-\phi_{i-1}\geq \phi_{i+1}-\phi_{i}$. As we will see in section \ref{SUquiverduality}, the condition in eq. \eqref{orderingcondition} is a consequence of the UV duality between the $[2]-SU(2)^N-[2]$ quiver theory and the $SU(N+1)+2(N+1)$F gauge theory, since the $i$-th CB parameter of the former can be related with the $N-i+1$-th Dynkin basis coordinates of the latter. 

The CFT prepotential is obtained from the perturbative one by a flop of an instanton of mass
\begin{equation}
M_I=\phi_1-\phi_2-\frac{1}{2}m_0^{(2)}.
\end{equation}
This is positive in the perturbative phase but becomes negative in the CFT phase, where the mass parameters satisfy the conditions $\phi_1\geq \phi_2$ and $\phi_1-\phi_2\gg |m_i|, |m^{(i)}_0|$. 

\begin{figure}[h]
\centering
\includegraphics[scale=0.20, trim={2cm, 1cm, 3cm, 2.3cm}, clip]{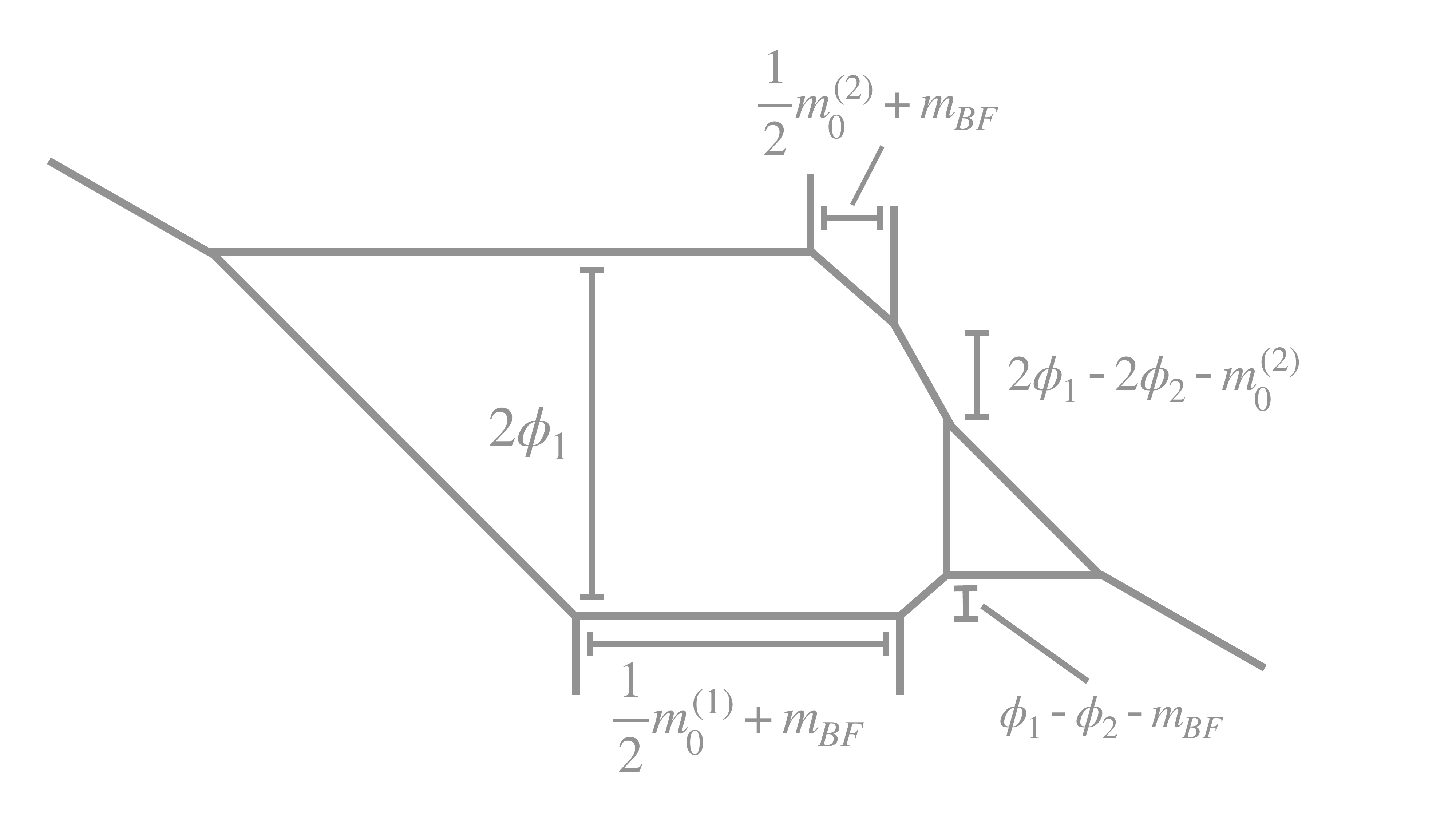}
\caption{The pq-web after the flop transition of the instanton $\phi_1-\phi_2-\frac{1}{2}m_0^{(2)}$.}\label{SU2SU2pastinfinity}
\end{figure}

The flop modifies the pq-web as in figure \ref{SU2SU2pastinfinity} and the CFT prepotential can be directly calculated from the tensions of the D3 branes wrapping the faces of the web.  
For later purposes, let us notice that all the representations of the global symmetry in eq. (\ref{X12prepot}) have a weight associated with a perturbative BPS state of the quiver theory. This tells us that the whole set of hypermultiplets can be obtained by applying the Weyl group to the perturbative bifundamentals or, in other words, that all representations are orbits of the Weyl group generated by acting on a bifundamental hypermultiplet. \\
\\
Having in mind the analysis above, we can now understand how to obtain the complete prepotential of $X_{1,2}$ directly from the perturbative one: 
\begin{itemize}
\item Order the CB parameter by enforcing the condition $\phi_1\geq \phi_2\geq 0$. 
\item Require the mass parameters to satisfy the CFT conditions $\phi_i \gg |m_i|, |m_0^{(i)}|$. 
\item Start from the pq-web in the perturbative phase and flop all edges that have negative length when the CFT conditions are enforced.
\item Obtain the invariant CB parameters from the monopole tensions in the CFT phase. Write the whole setup using the mass parameters associated with the global symmetry of the CFT.
\item Apply the Weyl group to the perturbative bifundamentals to obtain the hypermultiplet contribution and add it to $\mathcal{F}_{\text{CFT}}$.
\end{itemize}
Applying these steps to the pq-web in figure \ref{SU2SU2pastinfinity}, it is possible to obtain the complete prepotential of $X_{1,2}$ in eq. (\ref{X12prepot}). This procedure will be used in the next section to obtain the complete prepotential of the $X_{1,N}$ theory.

\section{Generalization: the \boldmath{$X_{1,N}$} complete prepotential}
\label{X1Nprep}
\subsection{CFT prepotential}
\label{CFTphase}
\begin{figure}[h]
\centering
\includegraphics[scale=0.22, trim={2cm, 1cm, 1cm, 2cm}, clip]{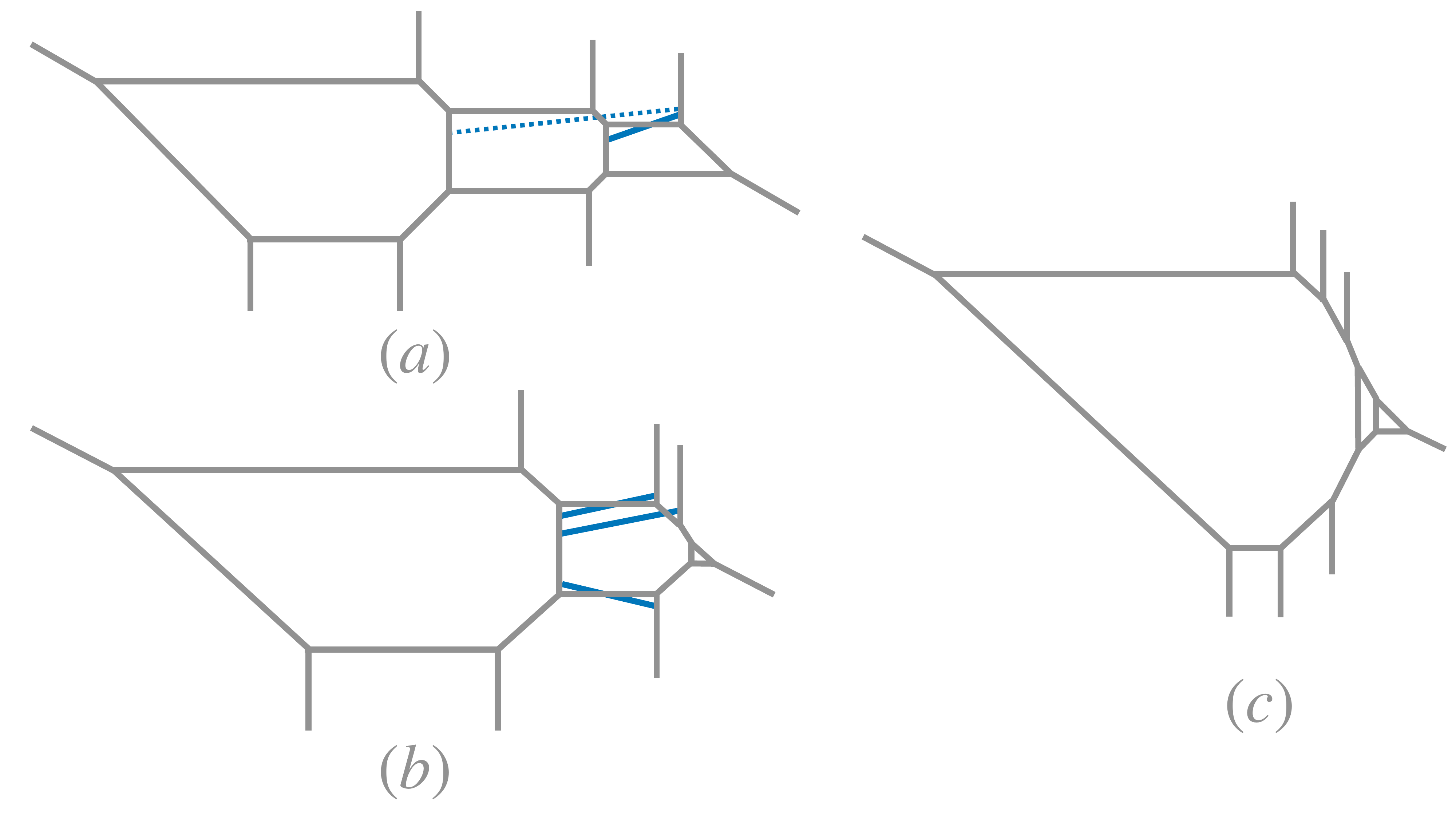}
\caption{Manipulation of $X_{1,3}$ pq-web via flopping instantons (in blue).}\label{WeylchamberX13}
\end{figure}

We have now what we need to generalize the previous analysis to the $X_{1,N}$ case. Let us start by writing the perturbative prepotential of the $SU(2)^N$  quiver theory 
\begin{equation}\label{pertX1N}
\mathcal{F}_{\text{pert.}}= \frac{4}{3}\sum_{i=1}^N \phi_i^3 + \frac{1}{2}\sum m_0^{(i)}\phi_i^2-\frac{1}{12}\sum_{i=1}^{N-1}|\phi_i\pm \phi_{i+1}\pm m_i|^3.
\end{equation}
Since we want to compute the prepotential in the CFT phase, we require the condition $\phi_i \gg |m_i|$ on the mass parameters together with the condition on the CB parameters $\phi_1\geq \phi_2\geq ...\geq \phi_N$. In this regime, the masses of the bifundamentals have a definite sign and the prepotential reduces to
\begin{equation}\label{pertprep}
\mathcal{F}_{\text{pert.}}= \frac{4}{3}\phi_N^3+\sum_{i=1}^{N-1}\phi_i^3 -\sum_{i=1}^{N-1}\phi_i\phi_{i+1}^2+\frac{1}{2}\sum_{i=1}^N m_0^{(i)} \phi_i^2 -\sum_{i=1}^{N-1} m_i^2\phi_i.
\end{equation}
The pq-web obtained by flopping the edges corresponding to these bifundamentals is depicted in figure \ref{WeylchamberX13}(a). On top of the perturbative bifundamental, we need to consider non-perturbative hypermultiplets that eventually have the sign of their mass flipped when going from the perturbative phase to the CFT one. Due to this condition, some additional lengths of the pq-web need to be flopped to reach the CFT phase. This generalizes the procedure for $X_{1,2}$ and it is still implemented by moving all the external NS5 branes to the first node of the quiver, as shown in figure \ref{WeylchamberX13}(b). 

Let us list the number of instantons that we need to flop to reach the CFT phase. Considering the $i$-th node with $i\neq N$, this has two NS5 branes associated with it and two instantons realized geometrically by D1 branes connecting the external NS5 branes with the internal one\footnote{Note that in the Weyl chamber $\phi_1\geq \phi_2\geq ...\geq \phi_N$, a generic internal node behaves like an $SU(2)$ theory with two flavors, which is UV completed by an $E_3$ SCFT, as can be seen from the pq-web in figure \ref{WeylchamberX13}(a). The instantons that we flop in the process are nothing but the two instantons of the $E_3$ theory.}, as shown in figure \ref{WeylchamberX13}(b). When the two branes are moved to the $(i-1)$-th node, the masses of the corresponding instantons
\begin{equation}
\phi_i-\phi_{i-1}+\frac{1}{2}m_0^{(i)}\pm m_i
\end{equation}
change sign in the operation and the prepotential changes accordingly. If $i=N$, we have a single instanton connecting the external NS5 brane with the internal one\footnote{The theory associated with the $N$-th behaves as an $\tilde{E}_1$ theory in this limit. The hypermultiplet that we flop is nothing but the instanton of this effective $\tilde{E}_1$ theory \cite{Aharony:1997bh}.}
\begin{equation}
\phi_N-\phi_{N-1}+\frac{1}{2}m_0^{(N)}.
\end{equation}
The NS5 brane that was moved to the $(i-1)$-th node has to be moved to the $(i-2)$-th face. This flops an additional instanton with mass
\begin{equation}
\phi_{i-1}-\phi_{i-2}+\frac{1}{2}m_0^{(i)}+\frac{1}{2}m_0^{(i-1)}
\end{equation}
if $i\neq N$ and 
\begin{equation}
\phi_{N-1}-\phi_{N-2}+\frac{1}{2}m_0^{(N-1)}+\frac{1}{2}m_0^{(N-2)}
\end{equation}
if $i=N$. Applying the previous formulas recursively, a total of $(N-1)^2$ instantons with masses
\begin{align}
&\phi_i-\phi_{i-1}+\frac{1}{2}\sum_{j=0}^{k-i} m_0^{(k-j)}\pm m_k, \,\,\, \forall i=2,...,k, \,\,\,\forall k=2,...,N-1,\\
&\phi_i-\phi_{i-1}+\frac{1}{2}\sum_{j=0}^{N-i} m_0^{(N-j)}, \,\, \forall i=2,...,N,
\end{align}
need to be flopped to reach the CFT phase (namely all the lengths of these edges are negative if we impose the CFT phase conditions). In terms of the $SU(2N)$ mass parameters, these can be written as
\begin{equation}
\phi_i-\phi_{i-1}-\frac{1}{2}y_{2i-1}-\frac{1}{2}y_{2i-2}+y_r, \,\,\, \forall i=2,...,[r/2], \,\,\,\forall  r=4,...,2N,
\end{equation}
so the perturbative prepotential of eq. (\ref{pertX1N}) changes as
\begin{equation}\label{cubicprep}
\mathcal{F}=\mathcal{F}_{\text{pert.}}+\frac{1}{6}\sum_{r=4}^{2N} \sum_{i=2}^{[r/2]} \left(\phi_i-\phi_{i-1}-\frac{1}{2}y_{2i-1}-\frac{1}{2}y_{2i-2}+y_r\right)^3.
\end{equation}
The effective couplings are obtained from the prepotential and read
\begin{align*}
&\frac{\partial^2 \mathcal{F}}{\partial \phi_j^2}=8\Phi_j-(2N-2j+3)\Phi_{j-1}+(2N-2j-1)\Phi_{j+1},\\
&\frac{\partial^2 \mathcal{F}}{\partial \phi_j\partial\phi_{j-1}}=-(2N-2j+3)\Phi_j+(2N-2j+1)\Phi_{j-1}
\end{align*}
for $j\neq 1,N$ and
\begin{align*}
&\frac{\partial^2 \mathcal{F}}{\partial \phi_1^2}= (2N-3)\Phi_2+(9-2N)\Phi_1,\,\,\,\,\,\frac{\partial^2 \mathcal{F}}{\partial \phi_1\partial \phi_2}= (2N-3)\Phi_1+(1-2N)\Phi_2,\\
&\frac{\partial^2\mathcal{F}}{\partial \phi_N^2}= 9\Phi_N-3\Phi_{N-1},\,\,\,\,\,\frac{\partial^2\mathcal{F}}{\partial \phi_N\partial \phi_{N-1}}= -3\Phi_N+\Phi_{N-1}
\end{align*}
if  $j=1,N$ where the following linear combinations
\begin{equation}
\Phi_i\equiv \phi_i-\frac{1}{2}\sum_{i=1}^{2i-1}y_i
\end{equation}
correspond to the invariant CB parameters. The cubic and quadratic terms of the complete prepotential are then fixed in order to reproduce the effective couplings.

Linear terms are fixed by reproducing the monopole tensions. From the previous requirements, we obtain the following CFT prepotential 
\begin{equation*}
6\mathcal{F}_{\text{CFT}}= 9\Phi_N^3+8\sum_{i=2}^{N-1}\Phi_i^3+(9-2N)\Phi_1^3-3\sum_{i=2}^{N} (2N-2i+3)\Phi_{i}^2\Phi_{i-1}+3\sum_{i=2}^{N} (2N-2i+1) \Phi_{i}\Phi_{i-1}^2-3\sum_i y_i^2 \Phi_1.
\end{equation*}
\subsection{Hypermultiplet contribution}
Having calculated the CFT prepotential of the $X_{1,N}$ theory, it remains to determine the spectrum of hypermultiplets that contribute to the complete prepotential of the theory. In the $N=2$ case, this was obtained in two separate ways: from the spectrum of the $[4]-SU(2)^2-[3]$ theory reduced after the decoupling of the additional seven flavors or by applying the Weyl group of $SU(4)$ to the perturbative hypermultiplet masses. In the following, we adopt the last technique. 

Let us consider a generic bifundamental stretching between the $i$-th and the $(i+1)$-th nodes. We can distinguish two sets of such bifundamentals, depending on their charge under the $(i+1)$-th node. The first set reads
\begin{equation}
\phi_i-\phi_{i+1}\pm m_i=\Bigg\{\begin{matrix}
\Phi_i- \Phi_{i+1} -y_{2i}\\
\Phi_i- \Phi_{i+1}-y_{2i+1}
\end{matrix},
\end{equation}
while the second 
\begin{equation}
\phi_i+\phi_{i+1}\pm m_i=\Bigg\{\begin{matrix}
\Phi_i+ \Phi_{i+1} +\sum_{k=1}^{2i-1} y_k +y_{2i+1}\\
\Phi_i+ \Phi_{i+1} +\sum_{k=1}^{2i-1} y_k +y_{2i}
\end{matrix}.
\end{equation}
The former set of bifundamentals belongs to the $\overline{\bold{2N}}$ antifundamental representation of $SU(2N)$ while the latter to the rank-$2i$ antisymmetric representation of $SU(2N)$ with dimension $\boldmath{\begin{pmatrix}
2N\\
2i
\end{pmatrix}}$. The action of the $SU(2N)$ Weyl group on these weights generates the following set of hypermultiplets
\begin{align}\label{reps}
\overline{\bold{2N}}: \,\,\,&\,\,\,\,\,\,\,\Phi_i-\Phi_{i+1}-y_j, \,\,\, i=1,...,N-1, \, j=1,...,2N,\\
\boldmath{\begin{pmatrix}
2N\\
2i
\end{pmatrix}}&:\,\,\, \Phi_i+\Phi_{i+1}+y_{j_1}+...+y_{j_{2i}}, \,\,\, i=1,...,N-1, \, 1\leq j_1<...<j_{2i}\leq 2N.\nonumber
\end{align}
We claim that the complete spectrum of (charged) hypermultiplets of the $X_{1,N}$ theory is fully determined by these representations. This conjecture can be tested by calculating the number of flops of the $X_{1,N}$ theory from the flops of the $X_{1,N-1}$ theory. In the $N=2,3,4$ cases, all possible flops of the pq-web were found to belong to the representations listed in eq. (\ref{reps}). Moreover, the same feature is common to quiver theories of rank-2, as noticed in \cite{Hayashi:2019jvx}. For these theories, all the hypermultiplets contributing to the prepotential can be obtained from the action of the Weyl group on perturbative hypermultiplets. This property seems common to quiver theories with flavors, while it is not necessarily enjoyed by other gauge theories UV completed by the same fixed points. In the latter case, some orbits of the Weyl group can consist only of non-perturbative hypermultiplets. This is the case of the UV fixed point enjoyed by $SU(3)_{\pm \frac{1}{2}}+$9 F and $[4]-SU(2)-SU(2)-[3]$. A rank-5 antisymmetric representation of the $SO(20)$ global symmetry group of the fixed point belongs to the BPS spectrum and, while in the quiver theory this can be obtained as an orbit of a bifundamental perturbative hypermultiplet, in the $SU(3)_{\pm \frac{1}{2}}$ theory the representation consists of non-perturbative states only. 

This conjecture is also compatible with the complete prepotential of the  $SU(N+1)_0+2(N+1)$F, which will be the main topic of the next section.

The complete prepotential of the theory then reads
\begin{align}\label{completeprepX1N}
&6\mathcal{F}_{X_{1,N}}=9\Phi_N^3+8\sum_{i=2}^{N-1}\Phi_i^3+(9-2N)\Phi_1^3-3\sum_{i=2}^{N} (2N-2i+3)\Phi_{i}^2\Phi_{i-1}+\\ \nonumber
&+3\sum_{i=2}^{N} (2N-2i+1) \Phi_{i}\Phi_{i-1}^2-3\sum_i y_i^2 \Phi_1+\\\nonumber
&+\sum_i\sum_{j}[|\Phi_j-\Phi_{j+1}-y_i|]^3+\\\nonumber
&+\sum_{k=1}^N\sum_{1\leq j_1<...<j_{2k}\leq 2N}[|\Phi_k+\Phi_{k+1}+y_{j_1}+y_{j_2}+...+y_{i_{2k}}|]^3.\nonumber
\end{align}
\section{\boldmath{$SU(N+1)_0+2(N+1)$}F theory and dualities}
\label{SUquiverduality}
Another independent check of the validity of the $X_{1,N}$ prepotential comes from dualities. Indeed, the prepotential of $X_{1,N}$ theory can be independently obtained by decoupling the flavors of the $[2]-SU(2)^N-[2]$ quiver theory, which is UV dual \cite{Mitev:2014jza} to $SU(N+1)_0+2(N+1)\,$F theory. Knowing the prepotential of the $SU(N+1)$ theory, together with the duality map, we will obtain the $X_{1,N}$ prepotential. Moreover, the duality will give us useful information about the CFT phase of the $X_{1,N}$ theory.

The UV fixed point of the $SU(N+1)_0+2(N+1)$F theory enjoys an $SU(2)\times SU(2N+2)\times SU(2)$ global symmetry\cite{Mitev:2014jza}.\footnote{Note that for $N=1$ the global symmetry actually enhances to $E_5=\text{Spin}(10)$.} Being this group a product of special unitary groups, we expect the mass parameters of the global symmetries to be related to distances between groups of branes of the same $(p,q)$ charges, as we will soon verify explicitly. 

We first start analyzing the simplest case, namely the $SU(3)_0+6$F theory and the construction of the duality map with $[2]-SU(2)^2-[2]$. Decoupling the four flavors, we first reach the $[2]-SU(2)^2$ theory, matching the calculation of section \ref{Completeprep}, and we then obtain the prepotential of $X_{1,2}$. Moreover, we calculate the complete prepotential of $SU(N+1)_0+2(N+1)$F and exploit the corresponding duality map with $[2]-SU(2)^N-[2]$. Finally, through decoupling, we obtain the complete prepotentials of $[2]-SU(2)^N$ and $SU(2)^N$, giving a further consistency check of the validity of the prepotential we found in eq. (\ref{completeprepX1N}).
\subsection{$SU(3)_0+6$\normalfont F} 
\begin{figure}[h!]
\centering
\includegraphics[scale=0.27, trim={0.2cm, 4cm, 0.2cm, 3cm}, clip]{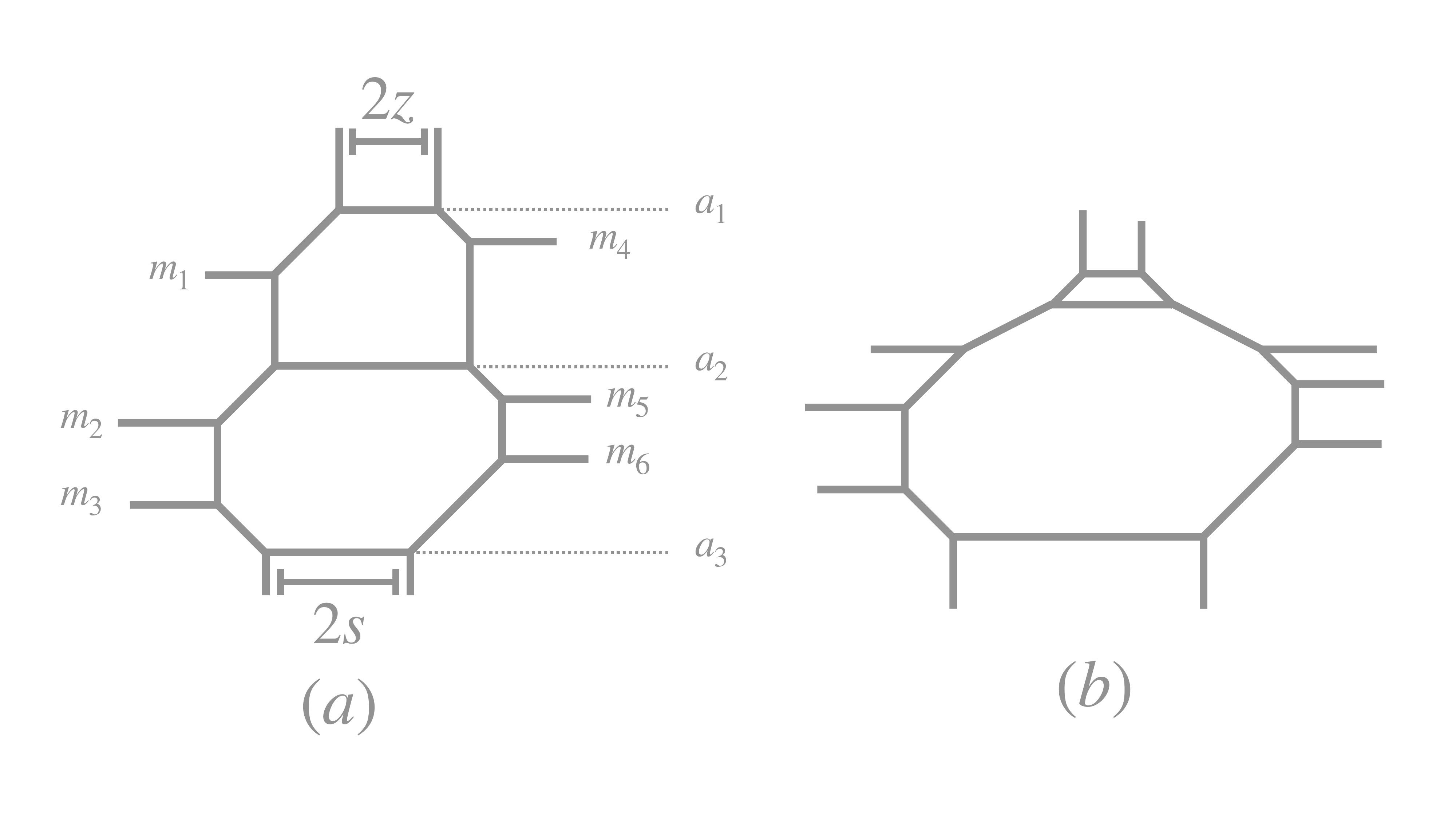}
\caption{$SU(3)_0+6$F (a) and in the CFT phase (b).}\label{SU36F}
\end{figure}

Let us begin by writing the perturbative prepotential of the $SU(3)_0+6$F theory. Its CB parameters $a_i.\,\,\,i=1,2,3$ obey the relation $\sum_i a_i=0$ and we choose a Weyl chamber $a_1\geq a_2\geq 0\geq a_3$. The perturbative prepotential reads
\begin{equation}\label{pertprep}
\mathcal{F}_{\text{pert.}}= \frac{1}{4}m_0(a_1^3+a_2^3+a_3^3)+\frac{1}{6} (a_1-a_2)^3+\frac{1}{6} (a_1-a_3)^3+\frac{1}{6} (a_2-a_3)^3-\frac{1}{12}\sum_{i=1}^3 \sum_{f=1}^6 |a_i-m_f|^3,
\end{equation}
where $m_f, \,\,\, f=1,...,6$ label the masses of the six flavors. The pq-web associated with the $SU(3)$ theory is shown in figure \ref{SU36F}(a). 
Before calculating the CFT prepotential, it is convenient to identify the mass terms associated with the global symmetry at the CFT point. The fixed point enjoys an $SU(6)\times SU(2)_z\times SU(2)_s$ global symmetry. The $SU(6)$ global symmetry is associated with the D5 branes of the pq-web, the $SU(2)_z$ with the NS5 branes in the upper side of the web, and the other $SU(2)_s$ with the NS5 on the lower side, see figure \ref{SU36F}. The mass parameters of $SU(6)$ correspond to the vertical positions of the D5 branes with respect to the center of mass, while the parameter of $SU(2)_s$ (resp. $SU(2)_z$) is identified with the horizontal distance between the upper (resp. lower) NS5 branes. 

The vertical coordinate of the COM of the D5 branes is $y_{	\text{COM}}= \frac{1}{6}\sum_{i=1}^6 m_i$, so the mass parameters read 
\begin{equation}\label{verticalpos}
y_i= \frac{5}{6}m_i-\frac{1}{6}\sum_{j\neq i} m_j,
\end{equation}
where $y_i$ labels the vertical position of the $i$-th brane, as shown in figure \ref{SU36F}(a). The parameter associated with the first $SU(2)_z$ group is $z=\frac{1}{2}x$, while the one associated with the $SU(2)_s$ group is $s=\frac{1}{2}y$.
Both parameters can be obtained by comparing the area of the two faces of the web with the derivatives of the perturbative prepotential and read
\begin{align}\label{horizontalpos}
&z=\frac{1}{2}x= \frac{1}{4}m_0+\frac{1}{4}\sum_{i=1}^6 m_i, \\
&s=\frac{1}{2}y= \frac{1}{4}m_0-\frac{1}{4} \sum_{i=1}^6 m_i.
\end{align}
In the CFT phase, the mass parameters are smaller than the CB ones, namely $a_i\gg |m_f|, \,\, i=1,2$ and $|a_3|\gg|m_f|$. This phase can be reached by flopping perturbative flavors only. The resulting phase of the pq-web is shown in figure \ref{SU36F}(b). For later purposes, it is convenient to adopt the conventions of \cite{Hayashi:2019jvx} and write the CB parameters $a_i$ in the Dynkin basis $\phi_1,\phi_2$ defined as $a_1=\phi_1, \,\,\, a_2=\phi_2-\phi_1, \,\,\, a_3=-\phi_2$. The effective couplings read
\begin{align}
&\frac{\partial^2 \mathcal{F}}{\partial \phi_1^2}=8\phi_1-4\phi_2+m_0+\sum_{i=1}^6 m_i,\\
&\frac{\partial^2 \mathcal{F}}{\partial \phi_1\partial\phi_2}=-4\phi_1+2\phi_2-\frac{1}{2}m_0-\frac{1}{2}\sum_{i=1}^6 m_i,\\
&\frac{\partial^2 \mathcal{F}}{\partial\phi_2^2}=2\phi_1+2\phi_2+m_0.
\end{align}
Correspondingly, the invariant CB parameters can be extracted from the effective coupling and read
\begin{equation}
\Phi_1=\phi_1+\frac{2}{3}z+\frac{1}{3}s, \,\,\, \Phi_2=\phi_2+\frac{1}{3}z+\frac{2}{3}s.
\end{equation}
Matching the monopole tensions, the CFT prepotential reads
\begin{equation}
6\mathcal{F}_{\text{CFT}}=8 \Phi_1^3+2\Phi_2^3 + 6\Phi_1 \Phi_2^2 - 
 12 \Phi_1^2 \Phi_2-6z^2\Phi_1-6s^2\Phi_2-3\sum_i y_i^2 \Phi_2
\end{equation}
and it is invariant under the Weyl reflections of the two $SU(2)$ groups acting as
$s\rightarrow -s$ and $z\rightarrow -z$ and under the Weyl group $S_6$ of $SU(6)$. \\
\\
To obtain the hypermultiplet contribution, let us first rewrite the flavor masses in terms of the invariant CB and mass parameters and act on the mass parameters with the Weyl group of $SU(6)\times SU(2)_z\times SU(2)_s$ global symmetry to construct the representations. The total contribution reads
\begin{equation}\label{partialrep}
\frac{1}{6}\sum_{i=1}^6[|\Phi_1\pm z-y_i|]^3+\frac{1}{6}\sum_{i=1}^6[|\Phi_2\pm s+y_i|]^3+\frac{1}{6}\sum_{i=1}^6[|\Phi_2-\Phi_1-y_i|]^3,
\end{equation}
where the first term describes the orbit generated from the perturbative flavors with masses $a_1-m_i$, the second the orbit generated from the flavors with masses $a_N+m_i$, and the third one generated from the remaining flavors.

\begin{figure}[h!]
\centering
\includegraphics[scale=0.24, trim={0.2cm, 0.2cm, 0.2cm, 0.2cm}, clip]{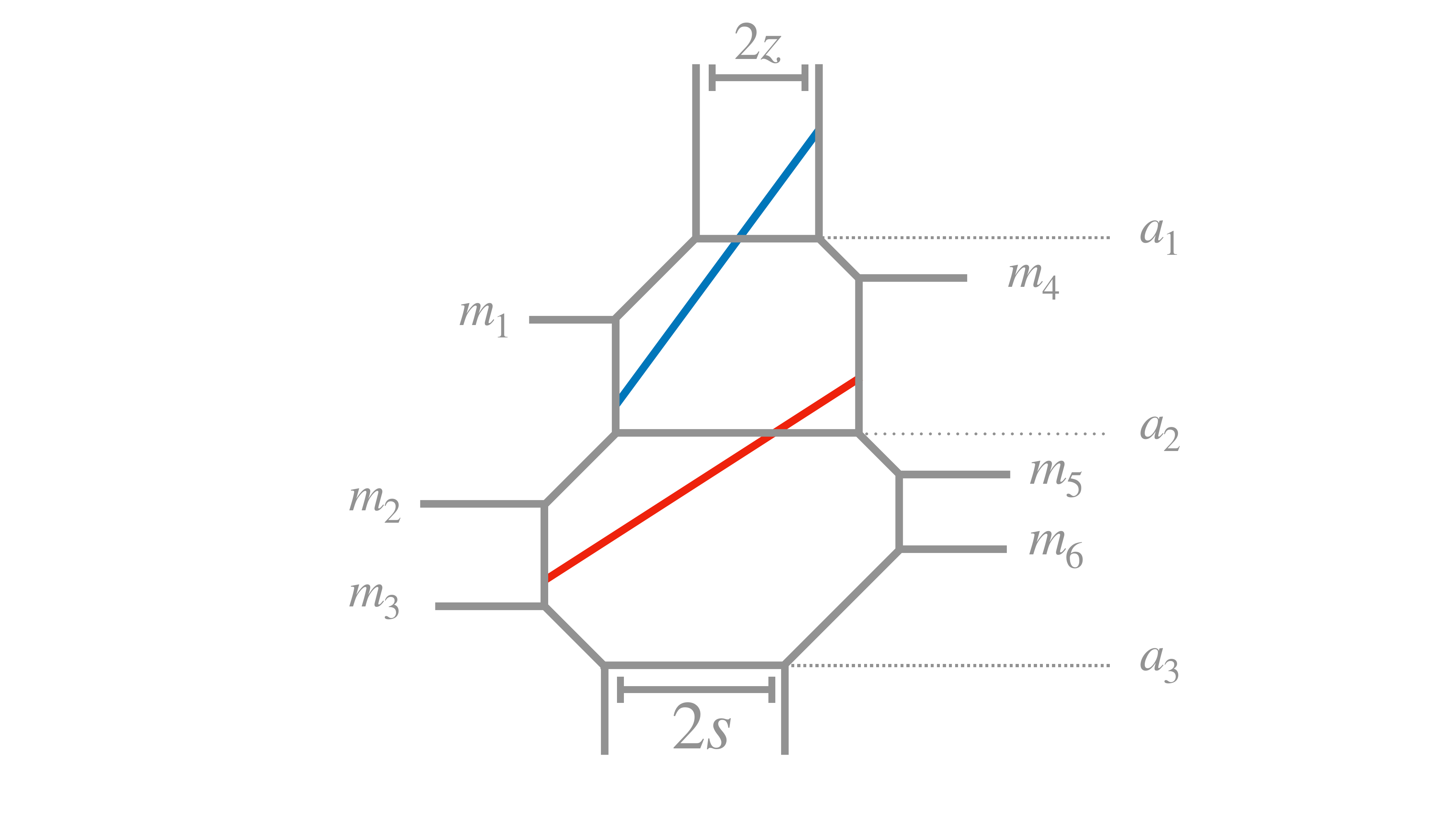}
\caption{$SU(3)_0+6$F missed hypermultiplet (red) and the additional instanton belonging to the $\bold{(\overline{6},2,1)}$ representation (blue).}\label{SU(3)6Fmissedhyper}
\end{figure}

\begin{figure}[h!]
\centering
\includegraphics[scale=0.23, trim={0.2cm, 0.3cm, 0.2cm, 0.3cm}, clip]{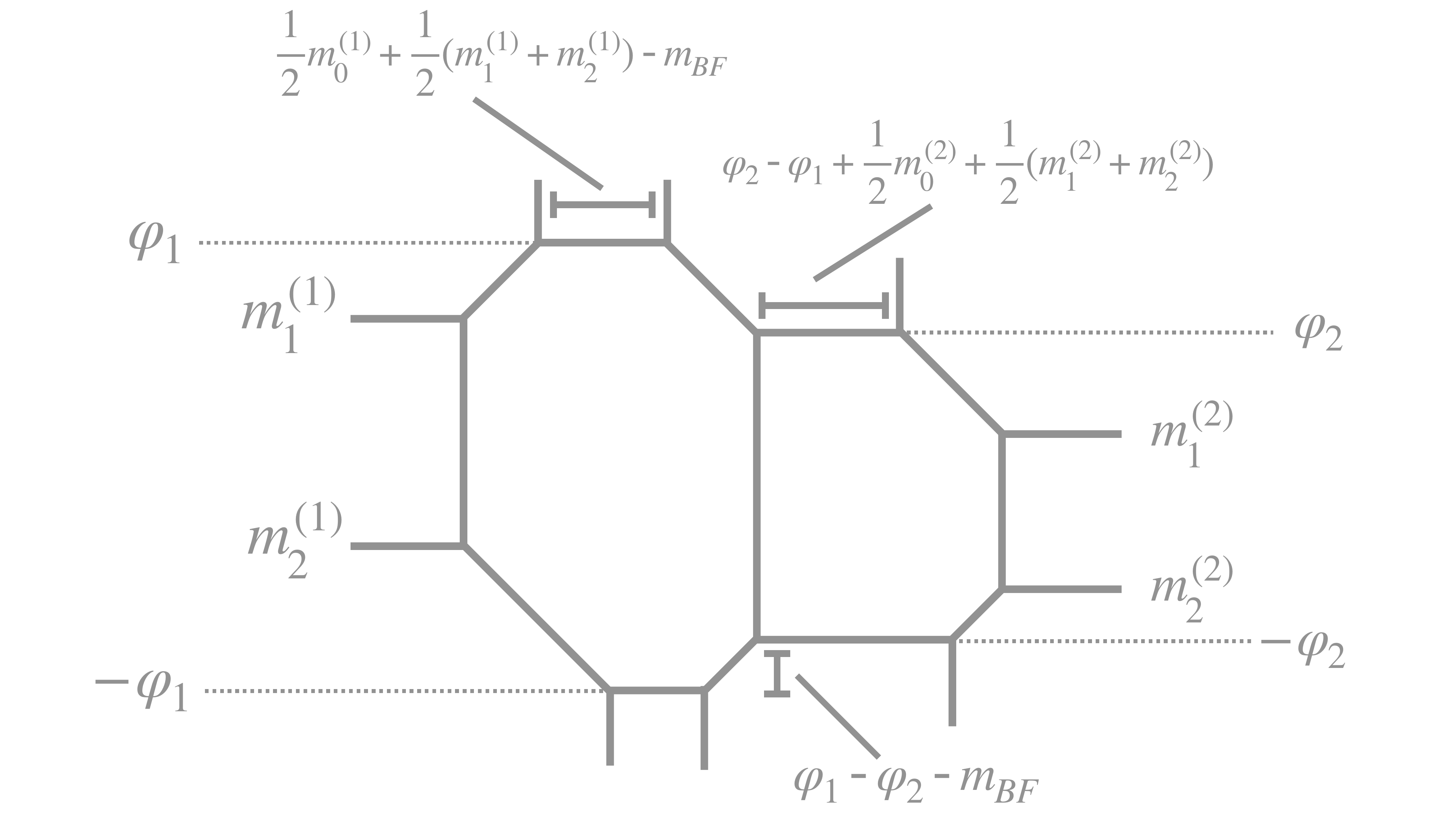}
\caption{$[2]-SU(2)-SU(2)-[2]$ quiver theory.}\label{2SU2SU22}
\end{figure}

As anticipated, not all the hypermultiplets belong to one of the representations listed in eq. (\ref{partialrep}). This is clear if we look at the pq-web: the red hypermultiplet in figure \ref{SU(3)6Fmissedhyper} does not belong to any of the representations in eq. (\ref{partialrep}). On the contrary, applying the Weyl reflection to its mass, the corresponding multiplet reads
\begin{equation}\label{additionalrep}
\Phi_1+\Phi_2-y_i-y_j-y_k, \,\, 1\leq i<j<k\leq 6
\end{equation}
namely it is a singlet under the two $SU(2)$s and a rank-3 antisymmetric representation of the remaining $SU(6)$ subgroup. All the states belonging to this representation are non-perturbative, so it is completely invisible at the perturbative level. 
We claim that the representations in eq. (\ref{partialrep}) together with the representations in eq. (\ref{additionalrep}) exhaust all the set of hypermultiplets of the $SU(3)_0$ theory. The complete prepotential reads then 
\begin{align}\label{completeprepSU(3)}
&6\mathcal{F}_{\text{compl.}}=8 \Phi_1^3+2\Phi_2^3 + 6\Phi_1 \Phi_2^2 - 
 12 \Phi_1^2 \Phi_2-6z^2\Phi_1-6s^2\Phi_2-3\sum_i y_i^2 \Phi_2+\\
&+\sum_i [|\Phi_1\pm z-y_i|]^3+\sum_i[|\Phi_2\pm s+y_i|]^3+\nonumber\\
&+\sum_i[|\Phi_2-\Phi_1-y_i|]^3+\sum_{1\leq i<j<k\leq 6}[|\Phi_1+\Phi_2-y_i-y_j-y_k|]^3.\nonumber
\end{align}
The $SU(3)_0$ theory is $S$-dual to an $[2]-SU(2)^2-[2]$ quiver theory. The duality map is obtained by looking at the pq-web of the quiver theory shown in figure \ref{2SU2SU22}. The mass parameters $y_i$ can be written in terms of the mass parameters $\varphi_i, m_0^{(i)}, m_j^{(i)}$ of the weakly coupled quiver description as
\begin{align*}
&y_1=\frac{1}{6}(m_0^{(1)}+2m_0^{(2)}+3 m_1^{(2)}+3m_2^{(2)}),\,\,\,\,\, y_2=\frac{1}{6}(m_0^{(1)}-m_0^{(2)})-m_{BF},\\
&y_3=\frac{1}{6}(-2m_0^{(1)}-m_0^{(2)}-3m_1^{(1)}-3m_2^{(1)}),\,\,\,\,\, y_4=\frac{1}{6}(m_0^{(1)}+2m_0^{(2)}-3 m_1^{(2)}-3m_2^{(2)}),\\
&y_5=\frac{1}{6}(m_0^{(1)}-m_0^{(2)})+m_{BF},\,\,\,\,\, y_6=\frac{1}{6}(-2m_0^{(1)}-m_0^{(2)}+3m_1^{(1)}+3m_2^{(1)}),\\
&s=\frac{1}{2}(m_1^{(1)}-m_2^{(1)}),\,\,\,\,\,z=\frac{1}{2}(m_1^{(2)}-m_2^{(2)}),\\
&\Phi_1=\varphi_2+\frac{1}{6}( m_0^{(1)} + 2 m_0^{(2)}),\,\,\,\Phi_2=\varphi_1+\frac{1}{6}(2 m_0^{(1)} + m_0^{(2)}).
\end{align*}
Several comments are now in order. The CFT phase of the $SU(3)$ theory can be reached by imposing the CFT conditions $a_i\gg |m_j|$ in the Weyl chamber $a_1\geq a_2\geq 0 \geq a_3$. In the quiver theory, these requirements translate into an ordering of the CB parameters associated with the two nodes $\varphi_1\geq \varphi_2$ and $\varphi_i \gg |m_i|$. In the $SU(3)$ theory, the CFT phase is reached from the perturbative phase by flopping flavors, while in the quiver theory, it is implemented by flopping instantons with masses
\begin{equation}
\varphi_2-\varphi_1+\frac{1}{2}m_0^{(2)}+\frac{1}{2}(m_1^{(2)}+m_2^{(2)}), \,\,\ \varphi_2-\varphi_1+\frac{1}{2}m_0^{(2)}-\frac{1}{2}(m_1^{(2)}+m_2^{(2)}).
\end{equation}
This generalizes the mechanism that was observed in the $X_{1,N}$ case and explains why we need to impose the Weyl chamber condition $\varphi_1\geq \varphi_2$ in the quiver theory to reach the CFT phase.

It is worth noticing that all the hypermultiplets of the quiver theory belong to orbits of the Weyl group containing perturbative states only. In particular, from the quiver perspective, instantons such the one in figure \ref{SU(3)6Fmissedhyper} belong to orbits generated from perturbative bifundamentals of the quiver theory, as well as the representations in eq. (\ref{partialrep}) are constructed as orbits of the Weyl group acting on the bifundamentals or the flavor hypermultiplets of the quiver theory. So, as conjectured in the $X_{1,N}$ case, also here all representations contain at least one perturbative state of the quiver theory.

Starting from the complete prepotential of this theory, we can decouple the four flavors to reach the $X_{1,2}$ theory. In the process, our prepotential in eq. (\ref{completeprepSU(3)}) reduces to the $X_{1,2}$ prepotential obtained in eq. (\ref{X12prepot}).

We are now ready to generalize the previous procedure to the $SU(N+1)_0$ case, which will be the main topic of the next section.
\subsection{$SU(M)+2M$\normalfont F}
Let us consider an $SU(M)$ theory with $2M$ flavors. As already remarked, the theory enjoys a UV fixed point at which the perturbative global symmetry enhances to $SU(2M)\times SU(2)_z\times SU(2)_s$. As for the $M=3$ case, this enhancement is manifest if we look at the pq-web construction of the theory: the $SU(2M)$ group is associated with the $2M$ D5 branes, while the $SU(2)$ groups are associated with the two couples of NS5 branes. The perturbative prepotential in the Weyl chamber $a_1\geq a_2\geq...\geq a_{M-1}\geq 0\geq a_M$ reads
\begin{equation}\label{prepSU(N)}
\mathcal{F}_{\text{pert.}}= \frac{1}{4}m_0\sum_{i=1}^Ma_i^2+ \frac{1}{6}\sum_{i<j} (a_i-a_j)^3-\frac{1}{12} \sum_{i=1}^{M}\sum_{j=1}^{2M} |a_i-m_j|^3.
\end{equation}
The mass parameters associated with the flavor symmetry can be obtained through the usual reasoning and read
\begin{equation}\label{y1SU(M)}
y_i= m_i -\frac{1}{2M} \sum_{i=1}^{2M} m_i,
\end{equation}
while the distance between the NS5 branes on the upper (resp. lower) side of the web is twice the mass parameter associated with the $SU(2)_z$ (resp. $SU(2)_s$) global symmetry and reads
\begin{align}\label{zs}
&2z=\frac{1}{2}m_0+\frac{1}{2}\sum_{i=1}^{2M}m_i,\\
&2s=\frac{1}{2}m_0-\frac{1}{2}\sum_{i=1}^{2M}m_i.\nonumber
\end{align}
The CFT phase is reached demanding $a_i\gg |m_j|,\,\, \forall i=1,...,M-1,\,\, j=1,...,2M$. The effective couplings in this phase read
\begin{align*}
&\frac{\partial^2\mathcal{F}}{\partial \phi_i^2}=2(i-1)\Phi_{i-1}+8\Phi_i-2(i+1)\Phi_{i+1},\,\,\,\, i\neq M-1\\
&\frac{\partial^2\mathcal{F}}{\partial^2\phi_{M-1}}=2(M-2)\Phi_{M-2}+(8-2M)\Phi_{M-1},\\
&\frac{\partial^2\mathcal{F}}{\partial \phi_i\partial\phi_{i+1}}=-2(i+1)\Phi_i+2i\Phi_{i+1},\,\,\,\, i\neq M-1\\
&\frac{\partial^2\mathcal{F}}{\partial\phi_{M-2}\partial\phi_{M-1}}=2(M-2)\Phi_{M-1}+2(1-M)\Phi_{M-2},
\end{align*}
where $\phi_i$ are the CB parameters in the Dynkin basis, defined from the usual $a_i$ coordinates as $a_1=\phi_1, \,\, a_i=\phi_i-\phi_{i-1}\,\,\,\forall i=2,...,M-1,\,\, a_M= -\phi_{M-1}$, and $\Phi_j$ represent the invariant CB parameters\footnote{These were already constructed in \cite{Mitev:2014jza}.}
\begin{equation}
\Phi_j= \phi_j+\frac{M-j}{M}z+\frac{j}{M}s, \,\,\, \forall j=1,...,M-1.
\end{equation}
Matching the expression for the monopole tensions, the CFT prepotential reads
\begin{equation*}
6\mathcal{F}_{\text{CFT}}= 8\sum_{i=1}^{M-2} \Phi_i^3+6\sum_{i=2}^{M-1} (i-1) \Phi_i^2\Phi_{i-1}-6\sum_{i=2}^{M-1} i\Phi_i\Phi_{i-1}^2 +(8-2M) \Phi_{M-1}^3-6z^2 \Phi_1-3\left(\sum_{i=1}^{2M} y_i^2+2s^2\right)\Phi_{M-1}.
\end{equation*}
Applying the Weyl symmetry to the masses of the perturbative flavors associated with the first and last nodes, we obtain the following representations
\begin{equation}
\Phi_1\pm z-y_j,\,\,\,\, \Phi_{M-1}\pm s +y_j, \,\,\, j=1,...,2M
\end{equation}
while applying it to the remaining flavors we obtain
\begin{equation}
\Phi_{i+1}-\Phi_i-y_j, \,\, i=1,...,M-2, \,\, j=1,...,2M. 
\end{equation}
These are respectrively the $\bold{(\overline{2M},2,1)}$, $\bold{(\overline{2M},1,2)}$ and $\bold{(\overline{2M},1,1)}$ representations of $SU(2M)\times SU(2)_z\times SU(2)_s$. 

As for the $M=3$ case, instantons connecting the $i$-th and the $(i+1)$-th internal NS5 branes belong to purely non-perturbative orbits
\begin{equation}
\Phi_i+\Phi_{i+1}-y_{j_1}-...-y_{j_{2i+1}}, \,\,\  1\leq k_1\leq ...\leq j_{2i+1}\leq 2M
\end{equation}
in the conjugate of the rank-$(2i+1)$ antisymmetric representation of $SU(2M)$. 

We conjecture that the complete prepotential reads
\begin{align}\label{prepSU(M)}
&6\mathcal{F}_{\text{compl.}}= 8\sum_{i=1}^{M-2} \Phi_i^3+6\sum_{i=1}^{M-1} (i-1)\Phi_i^2\Phi_{i-1}-6\sum_{i=1}^{M-1} i\Phi_i\Phi_{i-1}^2 +(8-2M) \Phi_{M-1}^3+\\ \nonumber
&-6z^2 \Phi_1-3\left(\sum_{i=1}^{2M} y_i^2+2s^2\right)\Phi_{M-1}+\\\nonumber
&+\sum_{i=1}^{2M}[|\Phi_1\pm z-y_i|]^3+\sum_{i=1}^{2M}[|\Phi_{M-1}\pm s+y_i|]^3+\\ \nonumber
&+\sum_{j=1}^{M-2}\sum_{i=1}^{2M}[|\Phi_{j+1}-\Phi_{j}-y_i|]^3+\sum_{j=1}^{M-2}\sum_{1\leq i_1<...<i_{2j+1}\leq 2M}[|\Phi_j+\Phi_{j+1}-y_{i_1}-...-y_{i_{2j+1}}|]^3.\nonumber
\end{align}
\subsection{Duality with $[2]-SU(2)^{M-1}-[2]$}
\begin{figure}[h!]
\centering
\includegraphics[scale=0.27, trim={0.1cm, 10cm, 0.2cm, 7cm}, clip]{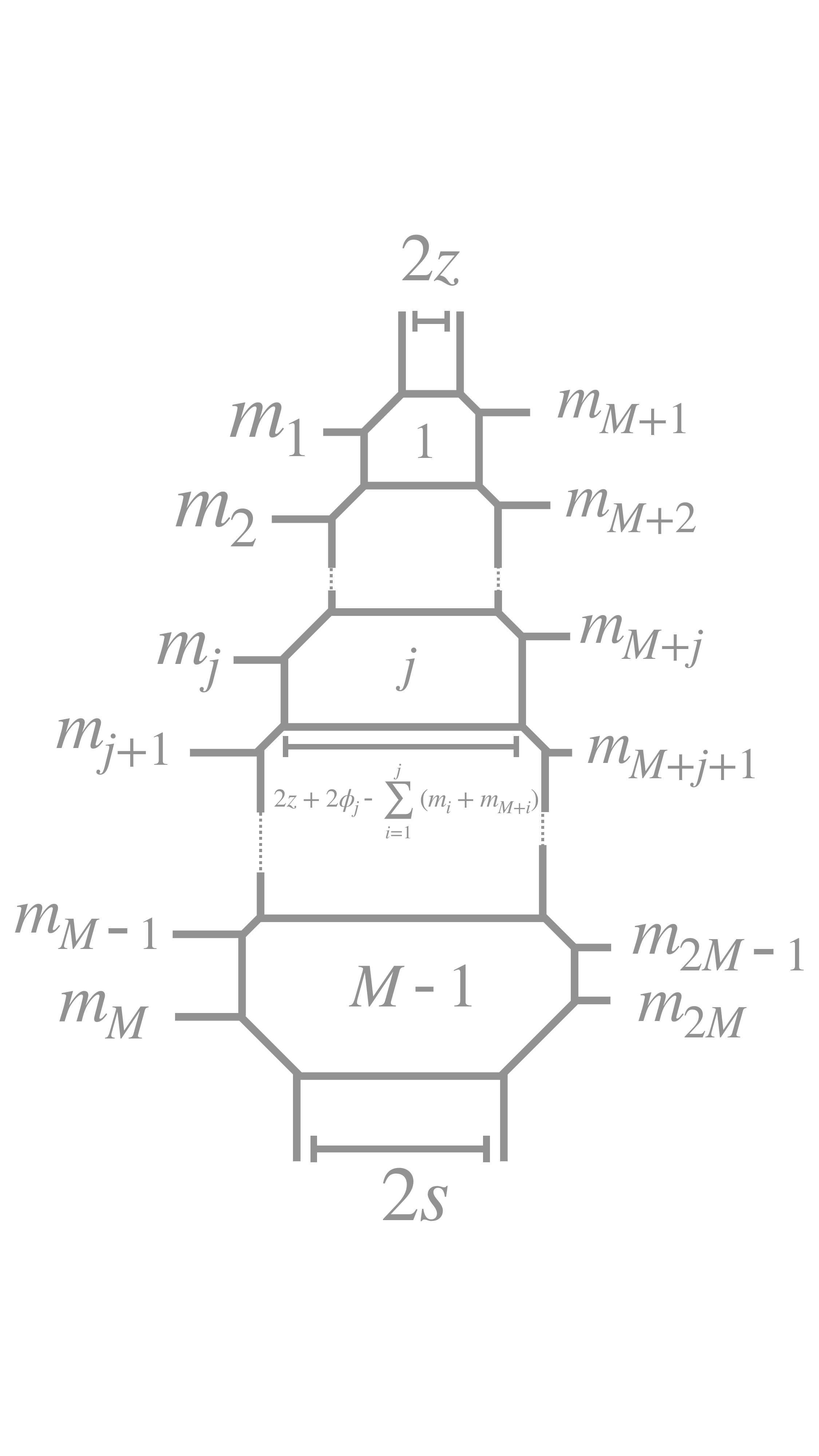}
\caption{$SU(M)+2M$ pq-web.}\label{dualitySU(M)}
\end{figure}

\begin{figure}[h!]
\centering
\includegraphics[scale=0.24, trim={0.2cm, 0.3cm, 0.2cm, 0.3cm}, clip]{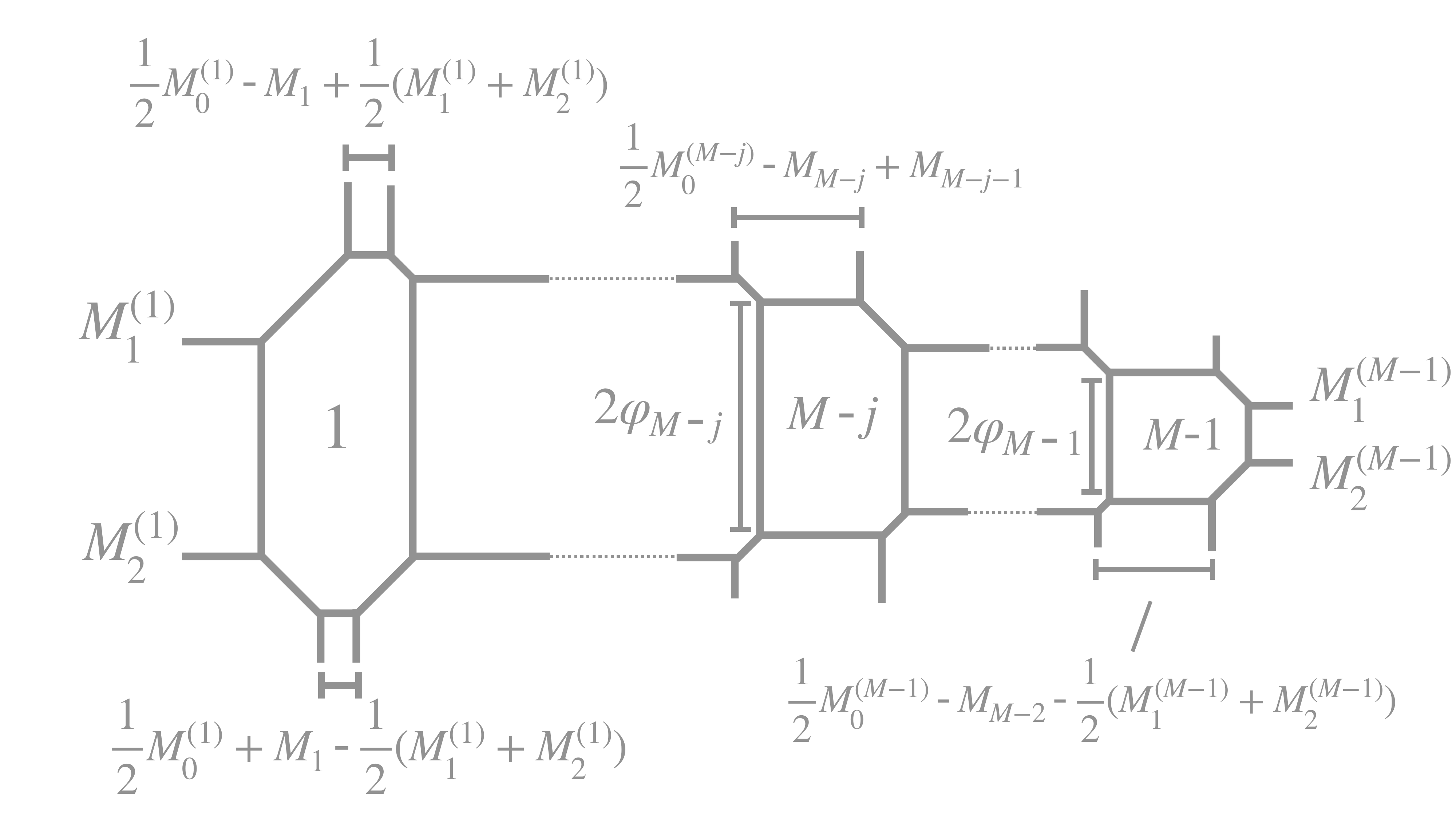}
\caption{$[2]-SU(2)^{M-1}-[2]$ pq-web.}\label{dualitySU(M)2}
\end{figure}

We can now exploit the duality between the $SU(M)+2M$F theory and $[2]-SU(2)^{M-1}-[2]$ and independently derive the $X_{1,M-1}$ prepotential. The map between the mass parameters of the two weakly coupled descriptions can be obtained from the corresponding pq-webs, which are related by an $S$-duality transformation. The two webs and their corresponding lengths are shown in figures \ref{dualitySU(M)} and \ref{dualitySU(M)2}.

In our conventions, the $j$-th face of the $SU(M)$ theory corresponds to the $(M-j)$-th face of the quiver. The mass parameters of the quiver theory are denoted by
\begin{equation}\label{massquiver}
M_j \,\,\,\,\forall j=1,...,M-1,\,\, M_0^{(j)}, \,\,M_1^{(1)}, \,\,M_1^{(M-1)},\,\, M_2^{(1)}, \,\,M_2^{(M-1)}, \,\,\varphi_j\,\,\,\forall j=1,...,M-1.
\end{equation}
Moreover, we will adopt the conventions of section \ref{X1Nprep}, relabeling the mass parameters in eqs. (\ref{zs}),\,(\ref{y1SU(M)}) as $y_i= \tilde{y}_{2M-2i+2}, \,\,\, i=1,...,M$ and $y_{M+i}= \tilde{y}_{2M-2i+1},\,\,\,i=1,...,M$. In these conventions, the CFT parameters read \cite{Mitev:2014jza}
\begin{align*}
&\tilde{y}_{1}= -\frac{1}{2M}\sum_{k=1}^{M-1} (M-k) M_0^{(k)}+\frac{1}{2}(M_1^{(1)}+M_2^{(1)}), \,\,\,\,\,\tilde{y}_{2}=-\frac{1}{2M}\sum_{k=1}^{M-1} (M-k) M_0^{(k)}-\frac{1}{2}(M_1^{(1)}+M_2^{(1)}),\\
&\tilde{y}_{n}=\frac{1}{2M}\sum_{k=1}^{[(n-1)/2]}kM_0^{(k)} -\frac{1}{2M}\sum_{k=[(n-1)/2]+1}^{M-1} (M-k) M_0^{(k)}+(-1)^{n+1}M_{[(n-1)/2]},\\
&\tilde{y}_{2M-1}= \frac{1}{2M} \sum_{k=1}^{M-1}k M_0^{(k)} -\frac{1}{2}(M_1^{(M-1)}+M_2^{(M-1)}), \,\,\,\,\tilde{y}_{2M}= \frac{1}{2M} \sum_{k=1}^{M-1}k M_0^{(k)} +\frac{1}{2}(M_1^{(M-1)}+M_2^{(M-1)}),\\
&z= \frac{1}{2}(M_1^{(M-1)}-M_2^{(M-1)}), \,\,\,\,\,\, s= \frac{1}{2}(M_1^{(1)}-M_2^{(1)})
\end{align*}
while the invariant CB parameters are
\begin{equation}
\Phi_j=\varphi_{M-j}-\frac{1}{2}\sum_{k=1}^{2(M-j)}\tilde{y}_k, \,\,\, \forall j=1,...,M-1.
\end{equation}
We can finally obtain the $X_{1,M-1}$ prepotential by decoupling the four flavors of the quiver theory, in the same way as we did for the $[4]-SU(2)^2-[3]$ theory in section \ref{Completeprep}. In the following, we will only state the final result of the decoupling of each couple of flavors. 

The decoupling of the flavors associated with the first face leads to the complete prepotential\footnote{Here we relabeled $\tilde{y}_{2i}\rightarrow y_{2i-2}$ and $\tilde{y}_{2i+1}\rightarrow y_{2i+1}$ since the decoupling limit sends the coordinate $\tilde{y}_2$ to infinity.}
\begin{align}\label{[2]SU(2)N}
&6\mathcal{F}_{\text{compl.}}= 8\sum_{i=1}^{M-2} \Phi_i^3+6\sum_{i=2}^{M-1} (i-1)\Phi_i^2\Phi_{i-1}-6\sum_{i=2}^{M-1} i\Phi_i\Phi_{i-1}^2 +(10-2M) \Phi_{M-1}^3+\\
&-6z^2 \Phi_1-3\sum_{i=1}^{2M-1} y_i^2\Phi_{M-1}+\sum_{i=1}^{2M-1}[|\Phi_1\pm z-y_i|]^3+\nonumber\\
&+\sum_{j=1}^{M-2} \,\sum_{i=1}^{2M-1}[|\Phi_{j+1}-\Phi_{j}-y_i|]^3+\nonumber&\\
&+\sum_{1\leq i_1<...<i_{2j+1}\leq 2M-1}[|\Phi_j+\Phi_{j+1}-y_{i_1}-...-y_{i_{2j+1}}|]^3\nonumber
\end{align}
with mass and invariant CB parameters
\begin{align*}
&y_{1}= -\frac{1}{(2M-1)}\sum_{k=1}^{M-1} (M-k) M_0^{(k)},\\
&y_{n}=\frac{1}{2(2M-1)}\sum_{k=1}^{[n/2]}(2k-1)M_0^{(k)} -\frac{1}{(2M-1)}\sum_{k=[n/2]+1}^{M-1} (M-k) M_0^{(k)}+(-1)^{n+1}M_{[n/2]},\\
&y_{2M-1}= \frac{1}{2(2M-1)} \sum_{k=1}^{M-1}(2k-1) M_0^{(k)} -\frac{1}{2}(M_1^{(M-1)}+M_2^{(M-1)}),\\
&y_{2M-2}= \frac{1}{2(2M-1)} \sum_{k=1}^{M-1}(2k-1) M_0^{(k)} +\frac{1}{2}(M_1^{(M-1)}+M_2^{(M-1)}),\\
&z=\frac{1}{2}(M_1^{(M-1)}-M_2^{(M-1)}), \,\,\,\,\,\,\,\,\Phi_j=\varphi_{M-j}-\frac{1}{2}\sum_{k=1}^{2(M-j)-1}y_k.
\end{align*}
Decoupling the last couple of flavors, we reproduce the prepotential in eq. (\ref{completeprepX1N}) by taking $M=N-1$, giving a further consistency check for the validity of the $X_{1,N}$ prepotential. 
\section{Conclusions and outlook}
\label{conclusion}
In this paper, we constructed examples of complete prepotentials of five dimensional rank-$N$ theories. Our specific theories of interest, namely $X_{1,N}$ and $SU(N+1)_0+2(N+1)$F have special features: the global symmetry enhances at their UV fixed point to a direct product of special unitary groups, whose algebras are realized by strings stretching between 7-branes of the same type. This allows us to identify the mass parameters associated with the enhanced global symmetry as distances between branes of the corresponding pq-web.

The procedure to construct complete prepotentials for quiver theories was clarified and compared using two other approaches: firstly by obtaining the complete prepotential from the decoupling of known rank-$2$ theories, such as the $[3]-SU(2)-SU(2)-[4]$ quiver theory, and then by exploiting the UV duality with the $SU(N+1)_0+2(N+1)$F gauge theories. In both cases, we developed a prescription to reach the CFT phase of the quiver theory and a procedure to obtain the complete set of hypermultiplets associated with these theories was given. We claim that the formulas in eqs. (\ref{completeprepX1N}) and (\ref{prepSU(M)}) describe the complete prepotentials of the $X_{1,N}$ and $SU(N+1)+2(N+1)$F theories. 

The large $N$ regime of $X_{1,N}$ was studied in \cite{Bertolini:2022osy}. In particular, both a supersymmetric and a SUSY breaking deformation of the fixed point were analyzed in the large $N$ limit of the web, unveiling the existence of a second order phase transition in the phase diagram of the theory for specific values of mass parameters. The construction of the complete prepotential contained in this paper can shed light on the phase diagram of these gauge theories. In particular, the CS terms for the background vector multiplet of the global symmetry, which are obtained by looking at the complete prepotential of the theory, can be used to label the various phases that the theory enjoys. Phases with different CS terms will then be separated by a phase transition. Some properties of these transitions can be inferred by looking at the jump of these invariants between the two phases. This aspect becomes particularly useful when supersymmetry is softly broken and we lack control over the non-perturbative dynamics of the theory, as was observed recently for the $E_1$ theory subject to a soft SUSY breaking deformation \cite{BenettiGenolini:2019zth}. It would be nice to perform a similar analysis on the $X_{1,N}$ theory subject to the deformations studied in \cite{Bertolini:2022osy} and to see also if, as happens for the $E_1$ theory in this context \cite{Bertolini:2021cew}, soft supersymmetry breaking deformations lead to instabilities on the Higgs branch of these theories. This is at the moment still work in progress. 

Both the $X_{1,N}$ and the $SU(N+1)$ theories can be seen as building blocks of more complicated CFTs, such as the $X_{M,N}$ and the $+_{M,N}$ theories \cite{Bergman:2018hin}. These theories can be mass deformed at low energies to quivers with large rank nodes and, in some cases, possess a gravity dual \cite{Gutperle:2018axv, Gutperle:2018vdd, Legramandi:2021uds}. As a consequence, a natural generalization of our analysis would be the calculation of complete prepotentials for these long quiver theories. The complete prepotential framework can then be used as a guideline to study deformations in the dual supergravity regime, both in the supersymmetric context \cite{Gutperle:2018axv} and when supersymmetry is broken.  

Finally, the possibility of generalizing the complete prepotential construction to other theories can be helpful to classify theories possibly admitting non-supersymmetric fixed points in their phase diagram when deformed by a non-supersymmetric deformation. A partial classification for fixed points of pure gauge theories was described in \cite{Akhond:2023vlb}, although we still lack a similar classification for theories with a global symmetry of arbitrary rank.
\subsection*{Acknowledgements}

We thank Mohammad Akhond, Oren Bergman, Matteo Bertolini, and Diego Rodriguez-Gomez for discussions. We are grateful to Oren Bergman and Matteo Bertolini 
 for very useful feedback on a first draft version. This work is partially supported by the Israel Science Foundation under grant No. 1254/22.
\addcontentsline{toc}{section}{Bibliography}

\bibliography{JHEP_v2bib}
\bibliographystyle{JHEP} 
\end{document}